\documentclass[twocolumn]{aastex61}
\received{}
\revised{}
\accepted{}
\submitjournal{ApJ}

\shorttitle{Updated BaSTI $\alpha$-enhanced models}
\shortauthors{Pietrinferni et al.}
\begin{document}

\title{The updated BaSTI stellar evolution models and isochrones: II. $\alpha-$enhanced calculations}

\correspondingauthor{Adriano Pietrinferni}
\email{adriano.pietrinferni@inaf.it}

\author{Adriano Pietrinferni}
\affiliation{INAF - Osservatorio Astronomico d'Abruzzo, Via M. Maggini, s/n, I-64100, Teramo, Italy}
%\nocollaboration

\author[0000-0002-0786-7307]{Sebastian Hidalgo}
\affil{Instituto de Astrof{\'i}sica de Canarias, Via Lactea s/n, La Laguna, Tenerife, Spain}
\affiliation{Department of Astrophysics, University of La Laguna, Via Lactea s/n, La Laguna, Tenerife, Spain}
%\nocollaboration

\author[0000-0001-5870-3735]{Santi Cassisi}
\affiliation{INAF - Osservatorio Astronomico d'Abruzzo, Via M. Maggini, s/n, I-64100, Teramo, Italy}
\affiliation{INFN, Sezione di Pisa, Largo Pontecorvo 3, 56127 Pisa, Italy}
%\nocollaboration

\author{Maurizio Salaris}
\affiliation{Astrophysics Research Institute, 
Liverpool John Moores University, IC2, Liverpool Science Park, 146 Brownlow Hill, Liverpool, L3 5RF, UK}
%\nocollaboration

\author{Alessandro Savino}
\affiliation{Astronomy Department, University of California, Berkeley, CA 94720, USA }
%\nocollaboration

\author{Alessio Mucciarelli}
\affiliation{Dipartimento di Fisica e Astronomia - Universit{\'a} degli Studi di Bologna, Via Piero Gobetti 93/2, I-40129, Bologna, Italy}
\affiliation{INAF - Osservatorio di Astrofisica e Scienza dello Spazio di Bologna, via Piero Gobetti 93/3 - I-40129, Bologna, Italy}
%\nocollaboration

\author{Kuldeep Verma}
\affiliation{Stellar Astrophysics Centre, Department of Physics and Astronomy, Aarhus University, Ny Munkegade 120, DK-8000 Aarhus C, Denmark}
%\nocollaboration

\author{Victor Silva Aguirre}
\affiliation{Stellar Astrophysics Centre, Department of Physics and Astronomy, Aarhus University, Ny Munkegade 120, DK-8000 Aarhus C, Denmark}
%\nocollaboration

\author{Antonio Aparicio}
\affiliation{Instituto de Astrof{\'i}sica de Canarias, Via Lactea s/n, La Laguna, Tenerife, Spain}
\affiliation{Department of Astrophysics, University of La Laguna, Via Lactea s/n, La Laguna, Tenerife, Spain}
%\nocollaboration

\author{Jason W. Ferguson}
\affiliation{Department of Physics, Wichita State University, Wichita, KS 67260-0032, USA}



\begin{abstract}
  This is the second paper of a series devoted to present an updated release of the BaSTI
  ({\it a Bag of Stellar Tracks and Isochrones}) stellar model and isochrone library.
  Following the publication of the updated solar scaled library,
  here we present the library for a $\alpha-$enhanced heavy element distribution.
  These new $\alpha-$enhanced models account for all improvements and updates in the reference solar metal
  distribution and physics inputs, as in the new solar scaled library.
  The models cover a mass range between 0.1 and $15~M_{\odot}$, 18 metallicities 
  between [Fe/H]=$-$3.20 and $+$0.06 with $[\alpha/Fe]=+0.4$ , and a 
  helium to metal enrichment ratio $\Delta{Y}$/$\Delta{Z}$=1.31. For each metallicity, He-enhanced stellar models
  are also provided.
  The isochrones cover (typically) an age range between 20~Myr and 14.5~Gyr, including 
  consistently the pre-main sequence phase. Asteroseismic properties of the theoretical models
  have also been calculated. Models and isochrones have been compared with results from independent
  calculations, with the previous BaSTI release, and also with selected observations,
  to test the accuracy/reliability of these new calculations. All stellar evolution tracks,
  asteroseismic properties and isochrones are made publicly available at http://basti-iac.oa-teramo.inaf.it
\end{abstract}

\keywords{galaxies: stellar content -- Galaxy: halo -- globular clusters: general -- stars: evolution, stars: general}

\section{Introduction} \label{sec:intro}

Accurate sets of stellar model calculations and isochrones
are necessary to interpret a vast array of spectroscopic and photometric
observations of individual stars, star clusters, and galaxies, both resolved and unresolved.  

Between 2004 and 2013  we have built and made publicly available   
the BaSTI ({\it a Bag of Stellar Tracks and Isochrones}) stellar models and isochrones library \citep{basti:04, basti:06, basti:07,
  basti:09, bastiwd, basti:13}\footnote{Available at: http://basti.oa-abruzzo.inaf.it/index.html.}, which 
has been extensively employed by the astronomical community.
This library covers a wide range of masses, evolutionary phases,
chemical compositions, and also provides integrated magnitudes and spectra of single-age, single-metallicity populations.

In the intervening years, improvements in the physics and chemical inputs of stellar model calculations 
have become available, most notably the revision of the solar metal composition \citep[e.g.,][and references therein]{bergemann:14}, plus new electron conduction opacities and some improved reaction rates. 
We have therefore embarked in an update of the BaSTI library, starting with models and isochrones for
solar scaled chemical compositions, presented in \cite{bastiac:18}.
In this new BaSTI release we have extended the mass range both towards lower and higher masses, and we also provide
some basic asteroseismic properties of the models.

This second paper presents the new BaSTI release of models and isochrones for
a $\alpha-$enhanced metal distribution, suitable to study populations in galactic haloes, spheroids and dwarf galaxies.
Our calculations are the latest addition to the list of $\alpha-$enhanced model sets computed over the last
22 years by various authors, sometimes restricted to the mass and metallicity range of Galactic halo stars
\citep{sw:98,don:00,salasnich:00,kim:02,dotter:08,don:12,parsecae}.

The plan of the paper is as follows. Section~\ref{code} briefly summarises the physics inputs
adopted in the new computations, and the heavy element distribution.
Section~\ref{library} presents the stellar model grid, including the
mass and chemical composition parameter space. 
Section~\ref{teocomp} shows comparisons between these new models and previous calculations available in the literature, 
whilst Sect.~\ref{obscomp} compares the models with selected observational benchmarks.
Final remarks follow in Sect.~\ref{conclusions}.

\section{Stellar evolution code, metal distribution and physics inputs}
\label{code}

We have employed the same stellar evolution code as in \cite{bastiac:18} --hereafter Paper~I-- and the reader
is referred to that paper for more information about 
the technical improvements since the first release of the BaSTI database.

The adopted $\alpha$-enhanced heavy element distribution is listed in Table~\ref{tab:mix}: 
The $\alpha$-elements O, Ne, Mg, Si, S, Ca, and Ti have been uniformly enhanced with respect to Fe
by [$\alpha$/Fe]$=+0.4$, compared to the \cite{caffau:11} solar metal distribution employed in Paper~I.
%\footnote{We remember that for the assumed scaled solar mixture, the actual solar metallicity is $Z_\odot = 0.0153$, while the corresponding actual ($Z/X$)$_\odot$ is equal to 0.0209.} adopted in \cite{bastiac:18}, i.e. the mixture by \cite{caffau:11}, supplemented when necessary by the abundances given by \cite{lodders:10}.
A uniform enhancement of all these $\alpha$-elements has been adopted also in other large stellar model grids 
\citep[see, e.g.,][]{kim:02, dotter:08, don:14} and is generally consistent with results from spectroscopy  
\citep[see, e.g.,][]{cayrel, hayes, mashonkina, melendez}. Just oxygen might be slightly 
more enhanced than the other $\alpha$-elements, by approximately an extra 0.1-0.15~dex.
An extra enhancement of oxygen 
makes isochrones of a given age and [Fe/H] slightly fainter and cooler in the main sequence turn-off region.
This effect is mimicked by considering a slightly older age for isochrones without the extra
oxygen, at the level of at most just 3-4\% percent if [O/Fe] is increased by 0.1-0.15~dex
\citep[see][]{dotter:07, don:12}.
Also, the $T_{\mathrm eff}$ of the lower main sequence in the regime of very low-mass stars would become slightly
cooler, at the level of about just 2\% \citep[see][]{dotter:07}.

The value [$\alpha$/Fe]$=+0.4$ adopted in our calculations
is close to the upper limits measured in the Galaxy; an interpolation in [$\alpha$/Fe]
between our solar scaled models and these new ones can provide accurate evolutionary tracks and isochrones
for any intermediate $\alpha$-enhancement, as verified with the 
DSEP model library, that includes several different values of [$\alpha$/Fe] between $-$0.2 and +0.8
\citep{dotter:08}.

Our adopted $\alpha$-enhanced distribution has been used consistently in the nuclear reaction 
network, in the calculation of radiative and electron conduction opacities,
as well as in the equation of state (EOS).
The sources for opacities and EOS are the same as described in Paper~I.

\startlongtable
\begin{deluxetable}{c|cc}
\tablecaption{The adopted $\alpha-$enhanced heavy element distribution.\label{tab:mix}}
\tablehead{
\colhead{Element} & \colhead{Number fraction} & \colhead{Mass fraction} }
\startdata
   C     &   0.132021 &       0.089896  \\
   N     &  0.030250   &     0.024020  \\
   O     &  0.603476    &    0.547368 \\ 
   Ne   &  0.123221      &  0.140962  \\
   Na   &  0.000850     &   0.001108  \\
   Mg   &  0.038070    &    0.052456  \\
   Al     & 0.001260      &  0.001927  \\
   Si    & 0.037210     &   0.059246  \\
   P     & 0.000120      &  0.000211  \\
   S     & 0.014810      &  0.026918  \\
   Cl    & 0.000070      &  0.000141  \\
   Ar    & 0.001380      &  0.003125  \\
   K     & 0.000050      &  0.000111  \\
   Ca   & 0.002240      &  0.005090  \\
   Ti     & 0.000090      &  0.000244  \\
   Cr    & 0.000200      & 0.000590  \\
   Mn   & 0.000130      &  0.000405  \\
   Fe    & 0.013830      &  0.043787  \\
    Ni    & 0.000720       &  0.002396  \\
\enddata
\end{deluxetable}

The nuclear reaction and neutrino energy loss rates, treatment of superadiabatic convection
(all calculations employ a mixing length ${\rm \alpha_{ML}=2.006}$ obtained from a solar model calibration),
outer boundary conditions, treatment of overshooting from the convective cores
and atomic diffusion (without radiative levitation), are all as described in detail in Paper~I.

Regarding the outer boundary conditions, for models with masses 
${\rm M\le0.45M_\odot}$ we use outer boundary conditions provided by the {\it PHOENIX}  non-gray
model atmospheres \citep[see,][and references therein]{bastiac:18} described in \citet{allard:12}.
More precisely, we employ the so-called {\it BT-Settl}
model set\footnote{This dataset is publicly available at  
http://phoenix.ens-lyon.fr/Grids/ }.

As in Paper~I, mass loss is included with the \cite{reimers} formula, and the free parameter $\eta$ is set to
0.3, following the asteroseismic constraints discussed in \cite{miglio:12}. 
  We continue to use the \cite{reimers} formula because its free parameter has been calibrated through
  asteroseismology
  in nearby open clusters, and also to be homogeneous with our solar scaled calculations.
  Prompted by our referee, we have calculated test models with masses equal to 0.8, 1.8
  and 4$M_{\odot}$ at a representative [Fe/H]=$-$1.2, employing the more
  modern \citep{cuntz} mass loss formula. We have implemented equation~4 in \citep{cuntz} with the free parameter
  $\eta_{SC}$ set to $8 \times 10^{-14}~M_{\odot}~yr^{-1}$ as recommended by the authors,
  multiplied by another free parameter $\eta_{1}$, as done in \citet{valcarce}. We have then 
  determined the value of $\eta_{1}$ that provides the best agreement with our calculations based on the Reimers
  formula. 
  We found that for the 0.8 and 1.8 $M_{\odot}$ models a value $\eta_{1}$=0.6 matches
  our Reimers calculations in terms of both 
  HRD tracks and amount of mass lost. For the 4$M_{\odot}$ a value $\eta_{1}$=0.3 gives the best match.

  For the 10 and 15$M_{\odot}$ models we have experimented using the \citet{dejager} mass loss formula instead
  of the Reimers one.
  Main sequence HRD and lifetimes are identical for both masses and both mass loss choices, despite
  the fact that models
  calculated with the \citet{dejager} formula lose 0.1 and 0.3$M_{\odot}$ respectively, against 
  0.01 and 0.02$M_{\odot}$ for the reference Reimers calculations.
  The HRD and lifetimes of the following evolution are also barely affected by the choice of the mass
  loss, even though
  by the end of core He-burning the calculations with \citet{dejager} formula
  have lost 0.32 and 0.62$M_{\odot}$ respectively, against 
  0.03$M_{\odot}$ for both Reimers computations.

\startlongtable
\begin{deluxetable}{cccccc}
  \tablecaption{Initial values of the heavy element mass fraction $Z$, helium mass fraction $Y$,
    the corresponding [Fe/H] and [M/H], and the additional (enhanced) values of $Y$ at fixed $Z$, of our 
    model grid. \label{tab:zy}}
\tablehead{
\colhead{Z} & \colhead{Y} & \colhead{[Fe/H]} & \colhead{[M/H]} & \colhead{${\rm Y_{enh,1}}$} & \colhead{${\rm Y_{enh,2}}$   } }
\startdata
   0.0000199 &  0.247013   &   $-3.20$  &  $-2.90$ &  0.275 & 0.30 \\
   0.0000996 &  0.247117   &   $-2.50$  &  $-2.20$ &  0.275 & 0.30 \\   
   0.0001988  & 0.247247   &   $-2.20$  &  $-1.90$ &  0.275 & 0.30 \\   
   0.0003974  & 0.247506   &   $-1.90$  &  $-1.60$ &  0.275 & 0.30 \\
   0.0006275  & 0.247807   &   $-1.70$  &  $-1.40$ &  0.275 & 0.30 \\
   0.0008860  & 0.248146   &   $-1.55$  &  $-1.25$ &  0.275 & 0.30 \\
   0.0012500  & 0.248620   &   $-1.40$  &  $-1.10$ &  0.275 & 0.30 \\
   0.0015720  & 0.249040   &   $-1.30$  &  $-1.00$ &  0.275 & 0.30 \\
   0.0019750  & 0.249569   &   $-1.20$  &  $-0.90$ &  0.275 & 0.30 \\
   0.0027850  & 0.250628   &   $-1.05$  &  $-0.75$ &  0.275 & 0.30 \\ 
   0.0039200  & 0.252112   &   $-0.90$  &  $-0.60$ &  0.275 & 0.30 \\
   0.0061700  & 0.255054   &   $-0.70$  &  $-0.40$ &  0.275 & 0.30 \\
   0.0077300  & 0.257100   &   $-0.60$  &  $-0.30$ &  0.275 & 0.30 \\
   0.0120800  & 0.262790   &   $-0.40$  &  $-0.10$ &\nodata & 0.30 \\
   0.0150700  & 0.266905   &   $-0.30$  &   ~~0.00 &\nodata & 0.30 \\
   0.0187500  & 0.271502   &   $-0.20$  &   ~~0.10 &\nodata & 0.30 \\
   0.0242700  & 0.278717   &   $-0.08$  &   ~~0.22 &\nodata & 0.30 \\
   0.0325800  & 0.289584   &   ~~0.06   &   ~~0.36 &\nodata & 0.32 \\
\enddata
\end{deluxetable}

\section{The $\alpha$-enhanced model library}
\label{library}

This new BaSTI $\alpha$-enhanced model library includes
calculations for 18 values of the initial metallicity --a larger number than
in the previous BaSTI release \citep{basti:06}-- 
ranging from ${\rm Z\approx2\times10^{-5}}$ (${\rm [Fe/H]=-3.20}$) to ${\rm Z\approx0.033}$
(${\rm [Fe/H]=+0.06}$).
The initial values of $Y$ at a given $Z$ have been fixed assuming a primordial $Y$=0.247 and a
helium-enrichment ratio $\Delta{Y}/\Delta{Z}=1.31$, as discussed in Paper~I.

This $\alpha$-enhanced grid has been calculated for the same [Fe/H] values of the solar scaled one of
Paper~I, therefore at a given [Fe/H] the values of $Z$ are higher than for the solar scaled grid,
because of the different metal mixture.

An important difference with respect to the solar scaled grid of Paper~I is that this new 
$\alpha-$enhanced release includes multiple values of the initial He abundance, at a given $Z$.
The complete list of available chemical compositions is given in Table~\ref{tab:zy}.  

The purpose of these calculations with several initial He abundances
  is to study stellar populations in environments hosting He-rich stars, such as the Galactic bulge, elliptical
  galaxies, and also globular clusters. In the case of individual globular clusters, 
  He-enhanced stellar populations display specific patterns of variations
  of C, N, O, Na, Mg and Al with respect to the standard $\alpha$-enhanced composition of the helium
  normal component
  \citep[see, e.g.][for recent reviews]{bl18, gratrev:19, cs20}. As long as the sum C+N+O is unchanged 
  by these abundance patterns (which seems to be the case for most of the clusters)
  $\alpha$-enhanced calculations are appropriate to study globular clusters' multiple populations in the
  HRD, in optical and generally in infrared colour-magnitude-diagrams (CMDs)
  \citep[see, e.g.][and references therein]{swff, basti:09, cs20}.
  For CMDs involving wavelengths shorter than the optical range, and for the very low-mass stars in the infrared,
  appropriate bolometric corrections that account for the specific metal abundance patterns
  need to be calculated and applied to our isochrones \citep[see][]{cs20}.

For each composition -- but for the He-enhanced ones (see below) --
we have computed 56 evolutionary sequences, in the mass range between $0.1~M_\odot$ and  $15~M_\odot$).
For initial masses below $0.2~M_\odot$ we computed evolutionary tracks
for masses equal to 0.10, 0.12, 0.15 and  $0.18~M_\odot$.
In the range between 0.2 and $0.7~M_\odot$ we employed a mass step equal to $0.05~M_\odot$.
Mass steps equal to $0.1~M_\odot$,  $0.2~M_\odot$, 0.5~$M_\odot$ and 1~$M_\odot$ have been
adopted for the mass ranges $0.7-2.6~M_\odot$, $2.6-3.0~M_\odot$,  $3.0-10.0~M_\odot$,
and masses larger than $10.0M_\odot$, respectively.
For the He-enhanced chemical compositions, the upper mass limit was set to 
$2.0M_\odot$, to cover the observed age range of the
  massive clusters that display multiple stellar populations, which has a lower limit around
$\sim1.2-1.4$~Gyr \citep[see, e.g.][and references therein]{cabrera20}  
  \footnote{We provide upon request helium-enhanced models more massive than $2M_\odot$.}.

All models less massive than $4.0~M_\odot$ have been computed from the pre-main sequence
(pre-MS)\footnote{We did not compute the pre-MS of models more massive than 4.0$M_{\odot}$
because their pre-MS timescale is well below the lowest possible age of our isochrones, that is dictated
by the total lifetime of the more massive models in our grid.},
whereas more massive model calculations started from the zero age MS.
Relevant to the pre-MS calculations,
the initial mass fractions of  D, ${\rm ^{3}He}$ and ${\rm ^{7}Li}$  are set
to $3.9\times10^{-5}$, $2.3\times10^{-5}$, and $2.6\times 10^{-9}$ respectively (see Paper~I).

As in Paper~I, all evolutionary models --but the very low-mass ones whose core H-burning lifetime is much
longer than the Hubble time--  have been calculated until the start of the thermal pulses 
on the asymptotic giant branch, or C-ignition for the more massive ones.
In case of the long-lived lower mass models, we have stopped the calculations
when the central H mass fraction is $\sim$0.3 (corresponding to ages already much 
larger than the Hubble time).

For each initial chemical composition we provide also an extended set of core He-burning models suited to study 
the horizontal branch (HB) in old stellar populations. For each pair ($Z$, $Y$) we have computed  
models of varying total mass (with small mass steps) and fixed He-core mass. Both He-core mass
and chemical abundances in the envelope of the HB models are taken from the model of a red giant branch (RGB) progenitor
at the He-flash, with an age of $\sim12.5$~Gyr. 

Prompted by the referee and the results by \citet{valcarce}, we have quantified the effect of changing the age
of the RGB progenitors by performing numerical experiments at [Fe/H]=$-$1.2.
A decrease of the progenitor age from the reference 12.5~Gyr to 6~Gyr lowers the He-core mass at helium ignition
by 0.004$M_{\odot}$ (from 0.4865$M_{\odot}$ to 0.4822$M_{\odot}$) but increases
the helium abundance in the envelope by $\Delta Y$=0.01
(from $Y$=0.26 to $Y$=0.27) due to the variation of the efficiency of the
first dredge-up. As a consequence, 
the luminosity of the zero age HB (ZAHB) is roughly unchanged, and all tracks for masses above
0.5$M_{\odot}$ are essentially identical to the case of 12.5~Gyr progenitors. 
Only HB models with mass below this threshold --unlikely to be found in 6~Gyr old stellar populations-- 
are affected by this large change of the progenitor age. These 
tracks are increasingly shifted to lower $T_{\mathrm eff}$ with decreasing mass, by up to 15\% for
the lowest HB mass (equal to 0.487$M_{\odot}$, with a
ZAHB effective temperature of 30,000~K for the model with a 12.5~Gyr
progenitor) but their ZAHB luminosity is unchanged.

When the age changes from 12.5 to 10~Gyr the variations of the He-core mass and surface helium abundance  
are about half these values, whilst an increase of the age from 12.5 to 14~Gyr leaves
core masses and helium abundances unaffected.

We have also performed a second test along the following lines. Our HB models are computed considering
He-core mass and envelope composition of a progenitor whose evolution is calculated with our reference choice
of mass loss efficiency --Reimers formula with $\eta$=0.3. This means that, for example, a HB model with total
mass equal to
0.5$M_{\odot}$ has been computed with core mass and envelope composition of a progenitor that at He ignition 
had a mass larger than this value.
To check whether this procedure introduces any systematics in our HB calculations, we have computed
the evolution of several 0.8$M_{\odot}$ RGB progenitor models at [Fe/H]=$-$1.2
(with an age at the tip of the RGB equal to about 12.5~Gyr) varying $\eta$ from 0 to 0.63, to reach 
masses between 0.8 and 0.487$M_{\odot}$ at the He-flash. 
We have then calculated the HB evolution of these masses, to compare with the corresponding results obtained with
our reference method to calculate HB models.
Also in this case, only HB models with mass below 0.5$M_{\odot}$ --with
very thin envelopes and inefficient H-burning shell-- are affected. In this mass range the He-core mass
has decreased
by 0.001$M_{\odot}$ compared to calculations for $\eta$=0.3, and the tracks are shifted to temperatures lower by at
most 7\% for the lowest mass. The ZAHB luminosities are unchanged. 

As for all evolutionary sequences available in the BaSTI library,
also these new tracks 
have been \lq{normalized}\rq to the same number of
points to calculate isochrones, and more in general for ease of interpolation and
implementation in stellar population synthesis tools.
As extensively discussed in \cite{basti:04} and Paper~I, this normalization is based on the
identification of some characteristic homologous points (key points) 
corresponding to well-defined evolutionary stages along each track.
The choice of the key points, and the number of points distributed between two consecutive key points 
are as described in Paper~I.
For each chemical composition, these normalized evolutionary tracks are used to compute extended sets of
isochrones for ages between 20~Myr and 14.5 Gyr (older
isochrones can be computed upon request from the authors). For the He-enhanced compositions the isochrone age range
is between $\sim600$~Myr and 14.5~Gyr.

The solar scaled calculations of Paper~I included four model grids, computed with different
choices regarding whether convective core overshooting, atomic diffusion and mass loss are included or
neglected in the calculations 
(see Table~3 in Paper~I).
For these $\alpha$-enhanced calculations we provide
just one grid, corresponding to what we consider to be a \emph{best physics scenario}, that corresponds to
\emph{Case a} of Table~3 in Paper~I. This means that these models all include convective core overshooting,
atomic diffusion and mass loss.

Bolometric luminosities and effective temperatures
along evolutionary tracks and isochrones have been 
translated to magnitudes and colours using sets of bolometric corrections (BCs) calculated as described in Paper~I.
More specifically, we calculated BCs with
the ATLAS~9 suite of programs \citep{kurucz70}, for the same $\alpha$-enhanced
metal distribution of the stellar evolution models.
As in Paper~I, these BCs have been complemented in the low $T_{\mathrm eff}$ and high-gravity
regime with the spectral library by \citet{Husser13} for [$\alpha$/Fe]=0.4,
calculated with the {\it PHOENIX} code \citep{Hauschildt99}.

Table~\ref{tab:photo} lists all photometric systems presently available in the library\footnote{Additional photometric systems can be added upon request from the authors.}, and provides all the relevant information
about the source for the response curve of each filter, and the zero-points calibration.

\startlongtable
\begin{deluxetable*}{llll}
  \tablecaption{The photometric systems presently available in the library.\label{tab:photo}}
\tablehead{
\colhead{Photometric system} & \colhead{Calibration} & \colhead{Passbands} & \colhead{Zero-points} }
\startdata
2MASS & Vegamag & \citet{Cohen03}  & \citet{Cohen03} \\
DECam & ABmag & CTIO \tablenotemark{a} & 0 \\
{\em Euclid} (VIS $+$ NISP) & ABmag & Euclid mission database\tablenotemark{b} & 0 \\
{\em Gaia} DR1& Vegamag & \citet{Jordi10}\tablenotemark{c}  &\citet{Jordi10} \\
{\em Gaia} DR2 & Vegamag &\citet{Maiz-Apellaniz18}(MAW)\tablenotemark{d}& MAW\\
{\em Gaia} DR3 & Vegamag & \citet{gaiadr3} & \citet{gaiadr3} \\
{\em GALEX} & ABmag & NASA\tablenotemark{e} & 0\\
{\em Hipparcos }+ {\em Tycho}&ABmag&\citet{Bessell12}&\citet{Bessell12}\\
{\em HST} (WFPC2) & Vegamag & SYNPHOT &SYNPHOT\\
{\em HST} ( WFC3) & Vegamag & HST User Documentation\tablenotemark{f} &WFC3 webpage\tablenotemark{g}\\
{\em HST} (ACS) & Vegamag & HST User Documentation\tablenotemark{f}  & ACS webpage\tablenotemark{h}\\
J-PLUS& ABmag& J-PLUS collab. \tablenotemark{f}& 0 \\
{\em JWST} (NIRCam) & Vegamag & JWST User Documentation \tablenotemark{g}&SYNPHOT\\
{\em JWST} (NIRISS) & Vegamag & JWST User Documentation \tablenotemark{g}&SYNPHOT\\
{\em Kepler} & ABmag&Kepler collab.\tablenotemark{h}&0\\
PanSTARSS1 & ABmag & \citet{Tonry12} & 0\\
SAGE & ABmag &  SAGE collab. & 0\\
Skymapper & ABmag & \citet{Bessell11}  & 0\\
Sloan & ABmag &  \citet{Doi10} &\citet{Dotter08}\\
{\em Spitzer} (IRAC) & Vegamag & NASA\tablenotemark{i} & \citet{Groenewegen06}\\
Str\"{o}mgren & Vegamag & \citet{Maiz-Apellaniz06}   &\citet{Maiz-Apellaniz06}\\
{\em Subaru} (HSC) & ABmag & HSC collab.\tablenotemark{j} & 0\\
{\em SWIFT} (UVOT) & Vegamag & NASA\tablenotemark{k} & \citet{Poole08}\\
{\em TESS}& ABmag & NASA\tablenotemark{l}& 0\\
UBVRIJHKLM & Vegamag & \citet{Bessel88,Bessel90}  &\citet{Bessel98}\\
{\em UVIT} (FUV+NUV+VIS)&ABmag&UVIT collab. \tablenotemark{m}&\citet{Tandon17}\\ 
Vera C. Rubin Obs. & ABmag & LSST collaboration\tablenotemark{n}& 0\\
VISTA & Vegamag &  ESO\tablenotemark{o} & \citet{Rubele12}\\
{\em WFIRST} (WFI) & Vegamag & WFIRST reference information\tablenotemark{p}& SYNPHOT\\
WISE & Vegamag & WISE collab.\tablenotemark{q} & \citet{Wright10}\\
\enddata
\tablenotetext{a}{\url{http://www.ctio.noao.edu/noao/node/13140}}
%\tablenotetext{b}{\url{http://euclid.esac.esa.int/epdb/db/SPV02/SPV02/EUC_MDB_MISSIONCONFIGURATION-SPV02_2018-06-16T140000.00Z_01.01.xml.html}}
\tablenotetext{b}{\url{https://www.cosmos.esa.int/web/euclid/home}}
\tablenotetext{c}{The nominal G passband curve has been corrected following the post-DR1 correction provided by \citet{Maiz-Apellaniz17}}
\tablenotetext{d}{Two different $G_{BP}$ passbands are provided for sources brighter and fainter than G=10.87, respectively.}
\tablenotetext{e}{\url{https://asd.gsfc.nasa.gov/archive/galex/Documents/PostLaunchResponseCurveData.html}}
\tablenotetext{f}{\url{https://hst-docs.stsci.edu/wfc3ihb/chapter-6-uvis-imaging-with-wfc3/6-10-photometric-calibration}}
\tablenotetext{g}{\url{https://www.stsci.edu/hst/instrumentation/wfc3/data-analysis/photometric-calibration/uvis-photometric-calibration}}
\tablenotetext{h}{\url{https://www.stsci.edu/hst/instrumentation/acs/data-analysis/zeropoints }}
\tablenotetext{i}{\url{http://www.j-plus.es/survey/instrumentation}}
\tablenotetext{j}{\url{https://jwst-docs.stsci.edu/}}
\tablenotetext{k}{\url{https://keplergo.arc.nasa.gov/CalibrationResponse.shtml}}
\tablenotetext{l}{\url{https://irsa.ipac.caltech.edu/data/SPITZER/docs/irac/calibrationfiles/spectralresponse/}}
\tablenotetext{m}{\url{https://hsc-release.mtk.nao.ac.jp/doc/index.php/survey/}}
\tablenotetext{n}{\url{https://heasarc.gsfc.nasa.gov/docs/heasarc/caldb/data/swift/uvota/index.html}}
\tablenotetext{o}{\url{https://heasarc.gsfc.nasa.gov/docs/tess/data/tess-response-function-v1.0.csv}}
\tablenotetext{p}{\url{https://uvit.iiap.res.in/Instrument/Filters}}
\tablenotetext{q}{\url{https://github.com/lsst/throughputs/tree/master/baseline}}
\tablenotetext{r}{\url{http://www.eso.org/sci/facilities/paranal/instruments/vircam/inst/}}
\tablenotetext{s}{\url{https://wfirst.gsfc.nasa.gov/science/WFIRST_Reference_Information.html}}
\tablenotetext{t}{\url{http://wise2.ipac.caltech.edu/docs/release/allsky/expsup/sec4\_4h.html\#WISEZMA}}
\end{deluxetable*}

Finally, adiabatic oscillation frequencies for p-modes for all models have been computed by using the Aarhus adiabatic oscillation package \citep{oscicode} as described in Paper I. We do not calculate g-mode frequencies
  because they 
have limited applications due to the mode identification issue, and the computation is expensive. We provide radial, dipole, quadrupole,  and octupole p-mode frequencies for the models with central hydrogen mass fraction larger than $10^{-4}$, and only the radial mode frequencies for more evolved models. We note that non-radial modes can have mixed character in evolved models, i.e. they behave like p- and g-modes depending on the depth. Although mixed modes have been observed in subgiant branch and RGB stars, their analysis as well as the comparison with stellar models are  still challenging. We have also calculated the frequency of maximum power ($\nu_{\rm max}$), the large frequency separation for the radial mode  frequencies ($\Delta\nu_0$), and the asymptotic period spacing for the dipole mode frequencies ($\Delta P_1$).

\subsection{A note on atomic diffusion}

All currently available public stellar model libraries that include atomic diffusion (as our new calculations), do
account for the effect of pressure gradients (gravitational settling), temperature and chemical gradients, but neglect radiative
levitation.
As shown by \citet{turcotte}, radiative levitation does not have any major impact on the solar model, 
on the solar calibration of the mixing length and the
initial helium abundance of the Sun (see their Table~ 6), but it is expected to have a more relevant effect on
models with less massive convective envelopes, like low-mass metal poor models around the
main sequence turn off \citep[see][]{richard}.
To this purpose, Fig.~7 of \citet{richard} compares selected evolutionary properties of models
with 0.8$M_{\odot}$, and initial [Fe/H]=$-$2.31, with and without the inclusion of radiative
levitation. This comparison shows that the evolutionary track with radiative levitation is almost identical to the one
calculated with atomic diffusion without levitation. There is just a small difference in $T_{\mathrm eff}$ around the main
sequence turn off, the track with levitation being cooler by
less than 50~K: Luminosities and evolutionary timescales are identical. The major difference is the surface abundance of Fe, that is
enhanced compared to the initial value in the models with levitation, and severely depleted in the models without levitation.
In conclusion, evolutionary tracks and isochrones with atomic diffusion without radiative levitation should be a very good
approximation to calculations that include also the effect of radiative accelerations, apart from the values of
(at least some) surface chemical abundances.

An additional issue with atomic diffusion has emerged from spectroscopic observations of surface chemical abundances in
stars with thin (in mass) convective envelopes \citep[see, e.g. the review by][]{screview:17}.
These observations clearly show that the atomic diffusion efficiency (including radiative levitation)
in real stars is at least partially reduced 
compared to the predictions from theory \citep[see, e.g.][for the case of two Galactic globular clusters with
different initial metallicity]{korn07, mucc11}, even though this does not seem to be the case for the Sun.

These results point to a partial inhibition of diffusion from/into the convective envelopes caused
by some unspecified competing mechanism, that may be dependent on the mass size of the
surface convective regions. Nothing of course can be said about the efficiency of diffusion in the inner layers.
The same Fig.~7 of \citet{richard} shows the case of reducing the effect of diffusion from/into the envelope of
the same 0.8$M_{\odot}$, [Fe/H]=$-$2.31 calculations. The effect is mainly to make the tracks around the turn off
increasingly
hotter when diffusion gets progressively less efficient (and surface abundance variations smaller, compared to the
initial abundance values), but luminosities and evolutionary timescales are unaffected.
We found a similar result after calculations of some test low-mass, metal poor models, switching off diffusion just
from below the convective envelopes. Tracks and evolutionary timescales are identical to the case of full diffusion,
apart from a hotter $T_{\mathrm eff}$ around the turn off, which changes by up to 90-100~K.

\subsection{Comparison with solar scaled calculations}
\label{ssae}

\citet{salaris:93} have shown that $\alpha$-enhanced stellar evolution tracks and isochrones
can be well mimicked in the HRD and CMDs by solar scaled ones with the same total metallicity [M/H].
In their analysis they couldn't assess the effect of an $\alpha$-enhancement on the bolometric corrections,
and used the same solar scaled BCs also for their $\alpha$-enhanced calculations.
\citet{cassisi:04} investigated the effect of an $\alpha$-enhancement on BCs and colours, finding that
the good agreement between solar scaled and $\alpha$-enhanced isochrones with the same [M/H] is preserved
in $VI$ and infrared CMDs, but is less satisfactory in $BV$ and shorter wavelength CMDs.
Similar conclusions are found when considering the DSEP isochrones.

Here we have compared these new $\alpha$-enhanced isochrones with our \cite{bastiac:18} solar scaled ones, to
check whether previous results are confirmed.
Indeed we find that in the HRD \citet{salaris:93} results are confirmed for ages above about 1~Gyr, and across
the whole range of [M/H] of our calculations. The formula given in Eq.~3 of \citet{salaris:93} that relates 
the model [M/H] to [Fe/H] and [$\alpha$/Fe] is consistent with our new calculations at the level of 0.01~dex, 
despite the different reference solar metal distribution.

\begin{figure*}[ht!]
\begin{center}
\includegraphics[width=5.0in]{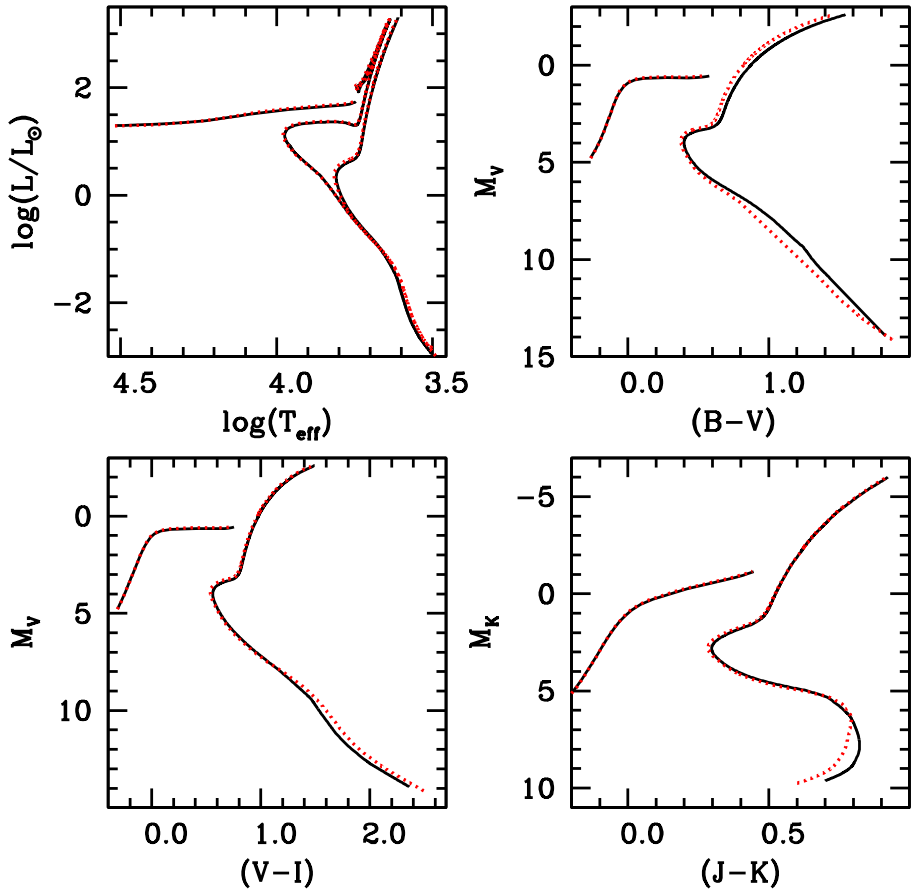}
\caption{{\sl Top-left panel:} HRD of solar scaled (solid lines) and $\alpha$-enhanced (dotted lines)
  isochrones for [M/H]=$-$1.40, $Y$=0.2478, and ages equal to 2 and 12~Gyr, respectively. The 12~Gyr isochrones
  are plotted up to the tip of the RGB, together with the corresponding ZAHBs.
  {\sl Top-right panel:} Optical $BV$ CMD of the 12~Gyr, [M/H]=$-$1.40 solar
  scaled (solid lines) and $\alpha$-enhanced (dotted lines) isochrones and ZAHBs. 
  {\sl Bottom-left panel:} Optical $VI$ CMD of the 12~Gyr, [M/H]=$-$1.40 solar
  scaled (solid lines) and $\alpha$-enhanced (dotted lines) isochrones and ZAHBs.
  {\sl Bottom-right panel:} Infrared $JK$ CMD of the 12~Gyr, [M/H]=$-$1.40 solar
  scaled (solid lines) and $\alpha$-enhanced (dotted lines) isochrones and ZAHBs.
  \label{fig:rescaling}}
\end{center}
\end{figure*}

Figure~\ref{fig:rescaling} shows a representative
comparison between selected $\alpha$-enhanced and
solar scaled isochrones from \cite{bastiac:18}, for [M/H]=$-$1.40 and $Y$=0.2478. These results
are farly independent of the chosen age (above 1~Gyr) and [M/H].

Differences in the HRD are at most equal to 1\% in $T_{\mathrm eff}$, and about 5\% in luminosity (0.02~dex)
around the TO and along the ZAHB ($\alpha$-enhanced isochrones being hotter and brighter).
This good agreement is preserved in the $VI$ and $JK$ CMDs, apart from the MS for masses below
about 0.5$M_{\odot}$, where the $\alpha$-enhanced
isochrones are systematically redder than the solar scaled ones 
by at most 0.07~mag in $(V-I)$, and bluer by at most 0.03~mag in $(J-K)$
(the lower MS wasn't explored by \citet{cassisi:04}).
The TO and ZAHB
magnitudes of the $\alpha$-enhanced isochrones are typically just about 0.04-0.05~mag brighter.
In the $BV$ CMD the $\alpha$-enhanced isochrones are systematically bluer by about 0.04~mag on average.

We also find that these colour differences increase when moving to CMDs involving shorter wavelength filters
like $U$, consistent with the results by \citet{cassisi:04}.

\section{Comparison with other model libraries}
\label{teocomp}

We compare here our new $\alpha$-enhanced isochrones to the previous $\alpha$-enhanced BaSTI release \citep[more specifically,
  the isochones calculated with
the Reimers parameter $\eta$=0.4][]{basti:06} and 
the DSEP \citep{dotter:08} $\alpha$-enhanced models \citep[comparisons of the previous BaSTI release
with earlier $\alpha$-enhanced model libraries are discussed in][]{basti:06}.
%
%publicly available,
%namely the previous BaSTI release \citep{basti:06}, DSEP \citep{dotter:08}, the Victoria model by \citet{don:12}, and the Yonsei-Yale one presented
%in \cite{kim:12}.
Comparisons are made in the theoretical HRD, to avoid additional 
differences introduced by the choice of the BCs, and we focus on old ages, typical
of $\alpha$-enhanced stellar populations.

\begin{figure}[ht!]
\begin{center}
\includegraphics[width=3.4in]{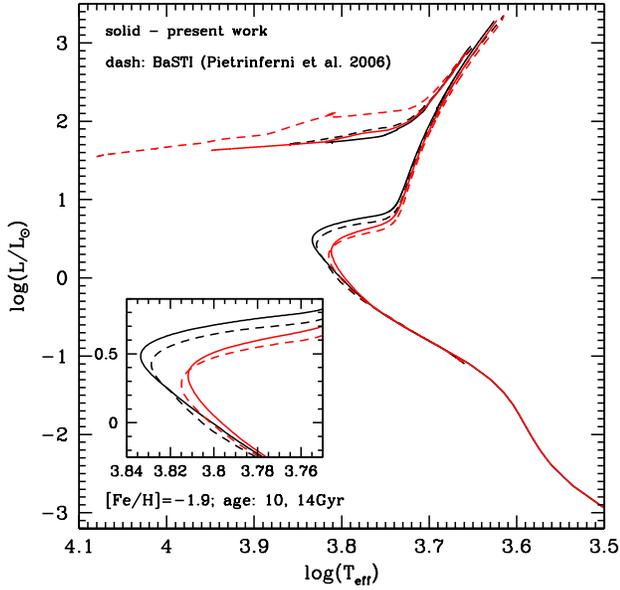}
\caption{HRD of 10 and 14~Gyr, [Fe/H]=$-1.9$ isochrones for our new calculations (solid lines) and the
  previous BaSTI $\alpha$-enhanced release \citep{basti:06}. The inset shows an enlargement of the MS turn off region.
  \label{fig:isom19}}
\end{center}
\end{figure}

\begin{figure}[ht!]
\begin{center}
\includegraphics[width=3.4in]{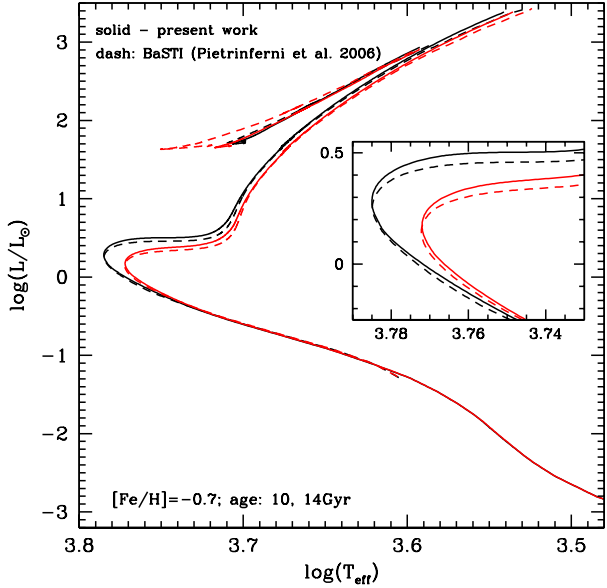}
\caption{As Fig.~\ref{fig:isom19} but for [Fe/H]=$-0.7$ isochrones.\label{fig:isom07}}
\end{center}
\end{figure}

Figures~\ref{fig:isom19} and ~\ref{fig:isom07} show the HRD of isochrones with ages equal to 10 and 14~Gyr and, respectively, 
[Fe/H]=$-1.9$ and [Fe/H]=$-0.7$, from both our new calculations and \cite{basti:06}.
At each [Fe/H] the initial $Y$ of the two sets of isochrones is the same within less than 1\%; 
the values of $Z$ are however much less similar, due to the different solar heavy element distributions.
At [Fe/H]=$-1.9$ our new models
have an initial $Z=0.0004$, compared to $Z=0.0006$ in \cite{basti:06} calculations, while at [Fe/H]=$-0.7$ the new calculations have
an initial $Z$=0.006, compared to $Z$=0.008 in the old BaSTI release.
Another difference between the two sets of isochrones arises from the inclusion of atomic diffusion in these new calculations,
which wasn't accounted for in \cite{basti:06} models, with the exception of the solar model computation to calibrate the
mixing length and the initial solar helium abundance
\citep[see,][for more details]{basti:04, basti:06}.

%\begin{figure}[ht!]
%\begin{center}
%\includegraphics[width=3.4in]{fig_iso_bastiae_fem13.eps}
%\caption{As Fig.~\ref{fig:isom19} but for an iron abundance equal to $[Fe/H]=-1.3$.\label{fig:isom13}}
%\end{center}
%\end{figure}

The MS section of the new isochrones \citep[extended to much lower masses compared to][]{basti:06}
is slightly hotter, due to the lower initial $Z$, but around the MS turn off (TO) the differences increase and
are metallicity- and age-dependent: This is due to the
combined effect of the inclusion of atomic diffusion in our new calculations, differences
in the nuclear cross sections for the H-burning discussed in \citet{bastiac:18}, and the different initial $Z$.
As a result, the isochrone TO luminosity is generally higher in these new calculations, whilst the
TO effective temperature is higher at the younger age and cooler at the older age displayed. These differences
decrease with increasing metallicity (a higher metallicity decreases the effect of atomic diffusion from the
convective envelopes, because of the more massive outer convective region).

As for the RGB, the new isochrones are systematically hotter than \citet{basti:06}, mainly due to the lower
initial $Z$, a result consistent
with the comparisons of the solar scaled isochrones \citep{bastiac:18}. The difference in $T_{\mathrm eff}$
increases with increasing [Fe/H]: It is on the order of 20-30~K at [Fe/H]=$-$1.9, increasing up to
$\sim$120~K at [Fe/H]=$-$0.7.

The core He-burning stage in the new isochrones is generally shifted to hotter $T_{\mathrm eff}$ and higher luminosities.
These differences are caused by the lower mass in \citet{basti:06} isochrones,
caused by a larger value of the Reimers parameter
$\eta$ ($\eta$=0.4 against $\eta=0.3$ in the new calculations) and
a brighter tip of the RGB (TRGB), that increases further the amount of mass lost along the RGB.

\begin{figure}[ht!]
\begin{center}
\includegraphics[width=3.4in]{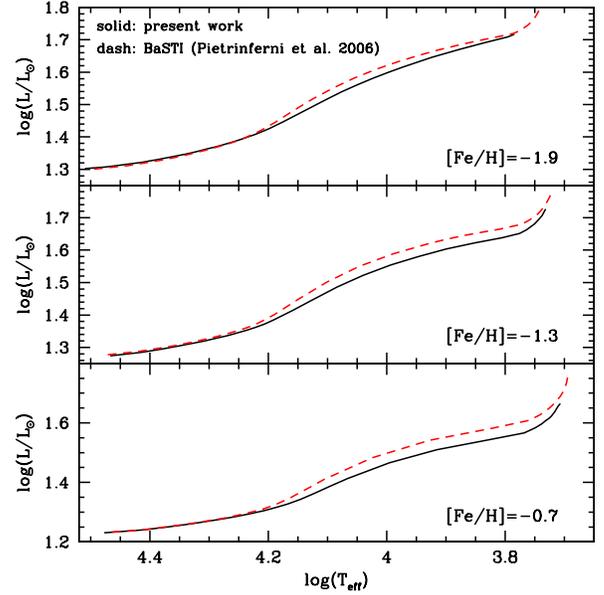}
\caption{HRD of ZAHB models from our new calculations and from \citet{basti:06} --solid and dashed lines, respectively-- for
  the three labelled values of [Fe/H]. The minimum and maximum mass of the 
  ZAHB models is the same in the two datasets (see text for more details).\label{fig:zahb}}
\end{center}
\end{figure}

\begin{figure*}[ht!]
\begin{center}
\includegraphics[width=6.5in]{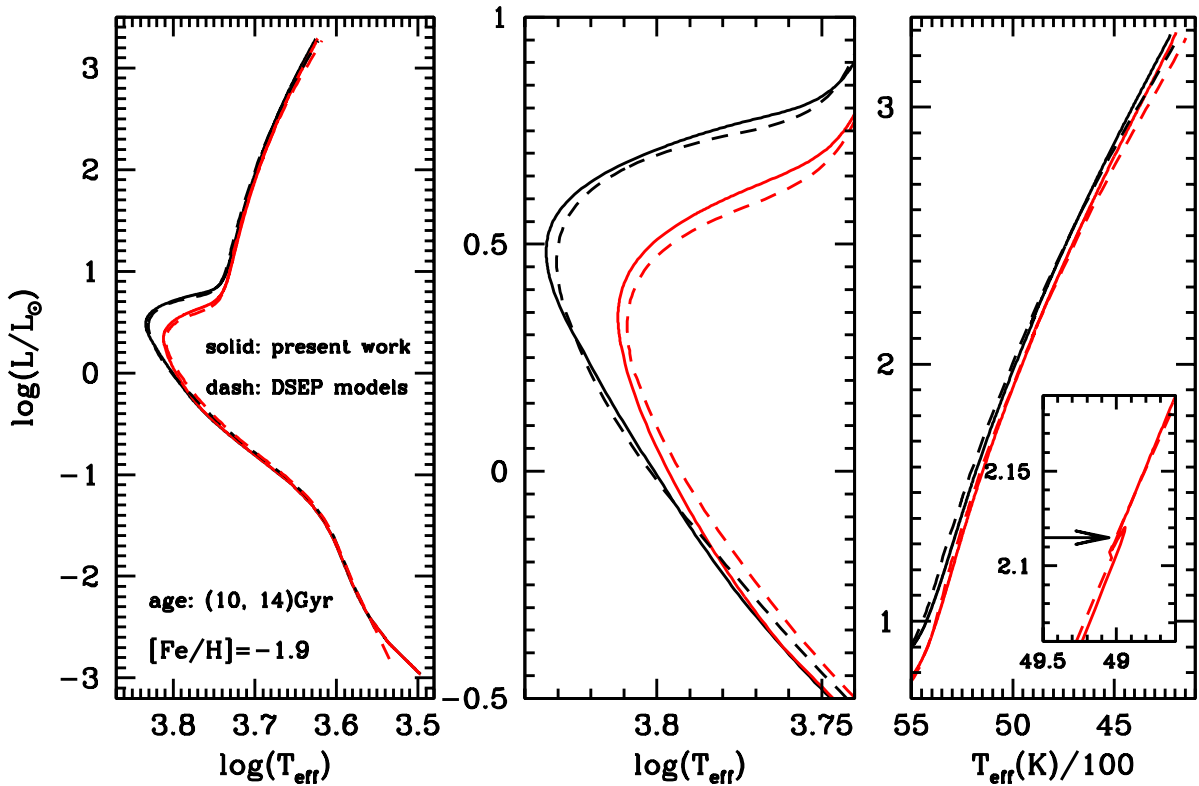}
\vskip -3.5truecm
\caption{\emph{Left panel:} HRD of 10 and 14~Gyr, [Fe/H]=$-1.9$ isochrones from the
  DSEP database, \citep[dashed lines --][]{dotter:08} and our new calculations (solid lines). \emph{Middle panel:} An enlargement
  of the TO region of the isochrones. \emph{Right panel:} Enlargement of the RGB portion of the isochrones. The inset shows
  the RGB bump region in the 14~Gyr isochrones; the arrow marks the location of the RGB bump
  in our calculations.\label{fig:dsepfe19}}
\end{center}
\end{figure*}

To explain more thoroughly the results of this comparison of core He-burning isochrones, it is
  helpful to study the corresponding ZAHB models. 
  Figure~\ref{fig:zahb} shows the HRD of ZAHB models (obtained from progenitors with an age of 12.5~Gyr at the TRGB)
  in our new calculations and \citet{basti:06}. A lower total mass shifts the ZAHB models towards hotter effective temperatures,
  hence lower
  luminosities, and this explains why the core He-burning \citet{basti:06} isochrones, with a lower
  evolving mass, are fainter than our new isochrones despite
  a generally brighter ZAHB at fixed $T_{\mathrm eff}$.

  To understand why the ZAHB luminosities of new and old calculations are different,
  three points need to be taken into account.
First, the new calculations include atomic diffusion, whose
impact on ZAHB models 
has been extensively investigated by \cite{cassisi:98} \citep[see also][and references
  therein]{cs:13}. Atomic diffusion increases the mass
of the He-core at He ignition for a given initial chemical composition,
and decreases the He abundance in the envelope.
The second point is that the improved electron conduction opacities employed in these new calculations
decrease the size of the He-core at He ignition, compared to \citet{basti:06} models \citep[for a detailed discussion on this
point we refer to][]{cassisi:07,bastiac:18}.
Finally, at a given [Fe/H], the new ZAHB models have a lower initial $Z$, because of the different solar metal distribution.

At the hot end of the ZAHB, above about 16,000~K,
it is the He-core mass that controls the luminosity, and here old and new calculations have roughly the same luminosity; this
happens because the increase of the He-core mass due to the inclusion of diffusion in the new models, 
approximately balances the decrease caused by the updated electron conduction opacities.
At lower effective temperatures the H-burning shell also contributes to the ZAHB luminosity, and the
situation is different: In this regime, despite the higher metallicity that tends to make the ZAHB fainter,
the old BaSTI calculations are systematically brighter than the new models, by about $\Delta\log(L/L_\odot)=0.02-0.04$~dex
at the level of the RR Lyrae instability strip.
This difference is driven by the 
reduction of the helium content of the envelope due to the inclusion of atomic diffusion (that decreases the
energy generation efficiency of the H-burning shell\footnote{This happens because the
  H-burning luminosity $L_H$  depends
  on the mean molecular weight $\mu$ as $L_H\propto\mu^7$. As a consequence of atomic diffusion He sinks during the
  MS, hence $\mu$ around the He-core and the H-burning luminosity both decrease
  \citep[see, e.g.][]{cs:13}.})
in the new models, together with 
the reduced He-core mass caused by the improved electron conduction opacities.

%\begin{figure*}[ht!]
%\begin{center}
%\includegraphics[width=6.5in]{fig_iso_dotter_fem13.eps}
%\vskip -3.5truecm
%\caption{As Fig.~\ref{fig:dsepfe19} but for a metallicity equal to $[Fe/H]=-1.3$.\label{fig:dsepfe13}}
%\end{center}
%\end{figure*}

The average mass of the new ZAHB models within the RR Lyrae instability 
strip ($M_{RR}$) taken at $\log(T_{\mathrm eff})=3.83$, is also different from \cite{basti:06} models.
At [Fe/H]=$-1.9$ our new models give $M_{RR}$=$0.675M_\odot$ 
compared to $0.72M_\odot$ in \cite{basti:06}, whilst
at [Fe/H]=$-1.3$, the new calculations provide $M_{RR}$=$0.635M_\odot$ against $\sim0.615M_\odot$ for the old
release. At [Fe/H]=$-0.7$ we get $M_{RR}=0.58M_\odot$, while $M_{RR}=0.568M_\odot$ in \cite{basti:06} models.

Next, we compare our new isochrones up to the TRGB with the DSEP ones \citep{dotter:08}, for the
same [Fe/H] and age values of the comparison with \cite{basti:06}.
These isochrones have been downloaded from the DSEP web tool, 2012
version, choosing the models with [$\alpha$/Fe]=+0.4, as in our calculations.
DSEP models also include atomic diffusion without radiative levitation, but 
there are differences in the physics inputs, most notably boundary conditions, electron conduction opacities,
and also the details of the EOS. Despite also a different reference solar heavy element distribution \citep{gs:98}
and a higher $\Delta{Y}/\Delta{Z}$, initial metallicities ($Z$=0.00041 and $Z$=0.0065)
and helium mass fractions ($Y$=0.2457 and $Y$=0.2555, respectively) are very close to our values
in these comparisons.

At [Fe/H]=$-1.9$ the two sets of isochrones are very similar, as shown in Fig.~\ref{fig:dsepfe19}.
The MS of DSEP calculations is slightly cooler, 
the TO luminosity is essentially the same, while the subgiant branch of our calculations is slightly brighter.
Along the RGB the DSEP isochrones have a different slope; they start hotter than ours at the base of the RGB, to become
cooler than our isochrones at higher luminosities, but $T_{\mathrm eff}$ differences are small, within $\pm$50~K. 
The inset in the right panel of Fig.~\ref{fig:dsepfe19} shows the RGB bump region along the 14~Gyr
old isochrones, that cannot be detected in the DSEP isochrone, likely due to the
sparse sampling of the RGB.

\begin{figure*}[ht!]
\begin{center}
\includegraphics[width=6.5in]{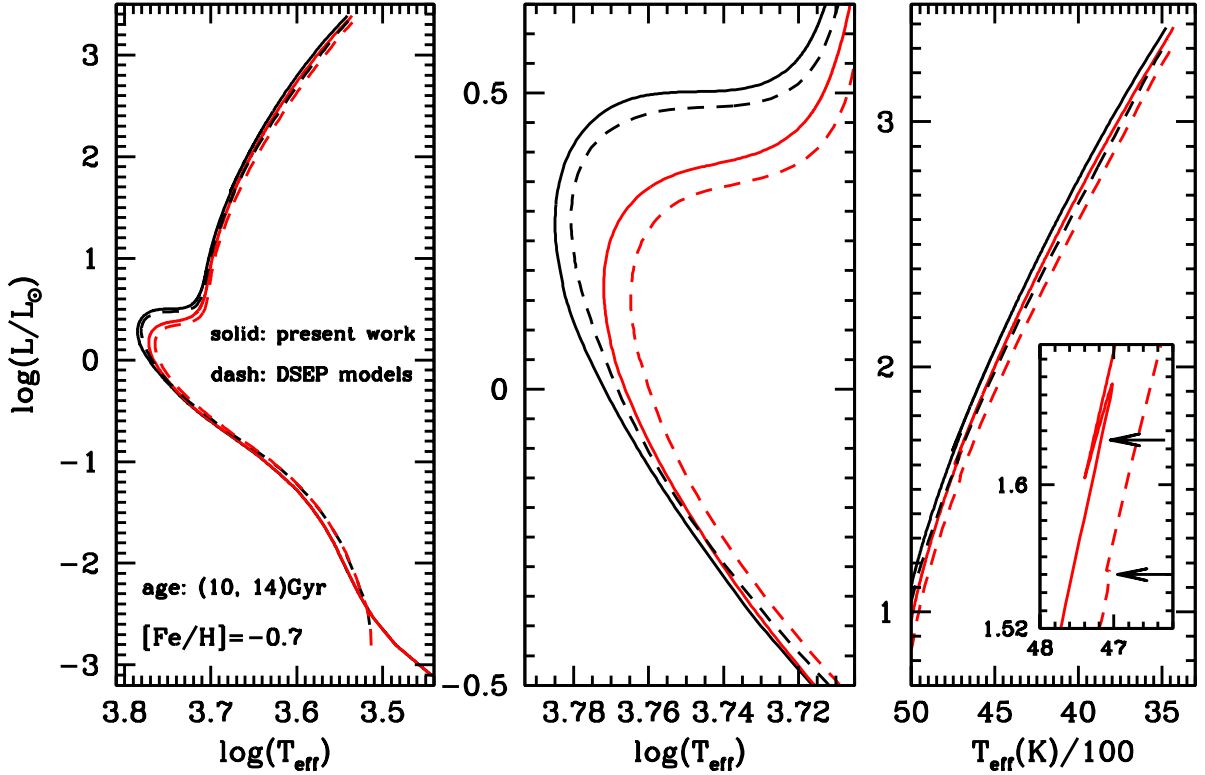}
\vskip -3.5truecm
\caption{As Fig.~\ref{fig:dsepfe19} but for [Fe/H]=$-0.7$. At this metallicity, the RGB bump can be identified
  also in the DSEP isochrone, and it is marked by an arrow.\label{fig:dsepfe07}}
\end{center}
\end{figure*}

The situation is similar at [Fe/H]=$-$0.7, as shown by Fig.\ref{fig:dsepfe07}, but the $T_{\mathrm eff}$
differences along the MS and at the TO are slightly amplified at this higher metallicity. On the RGB the 
DSEP isochrones are this time systematically cooler than ours, by $\sim 60-80$~K, and the RGB bump
in the 14~Gyr old isochrone is fainter than our results by $\Delta\log(L/L_\odot)\sim0.06$~dex.

\begin{figure}[ht!]
\begin{center}
\includegraphics[width=3.4in]{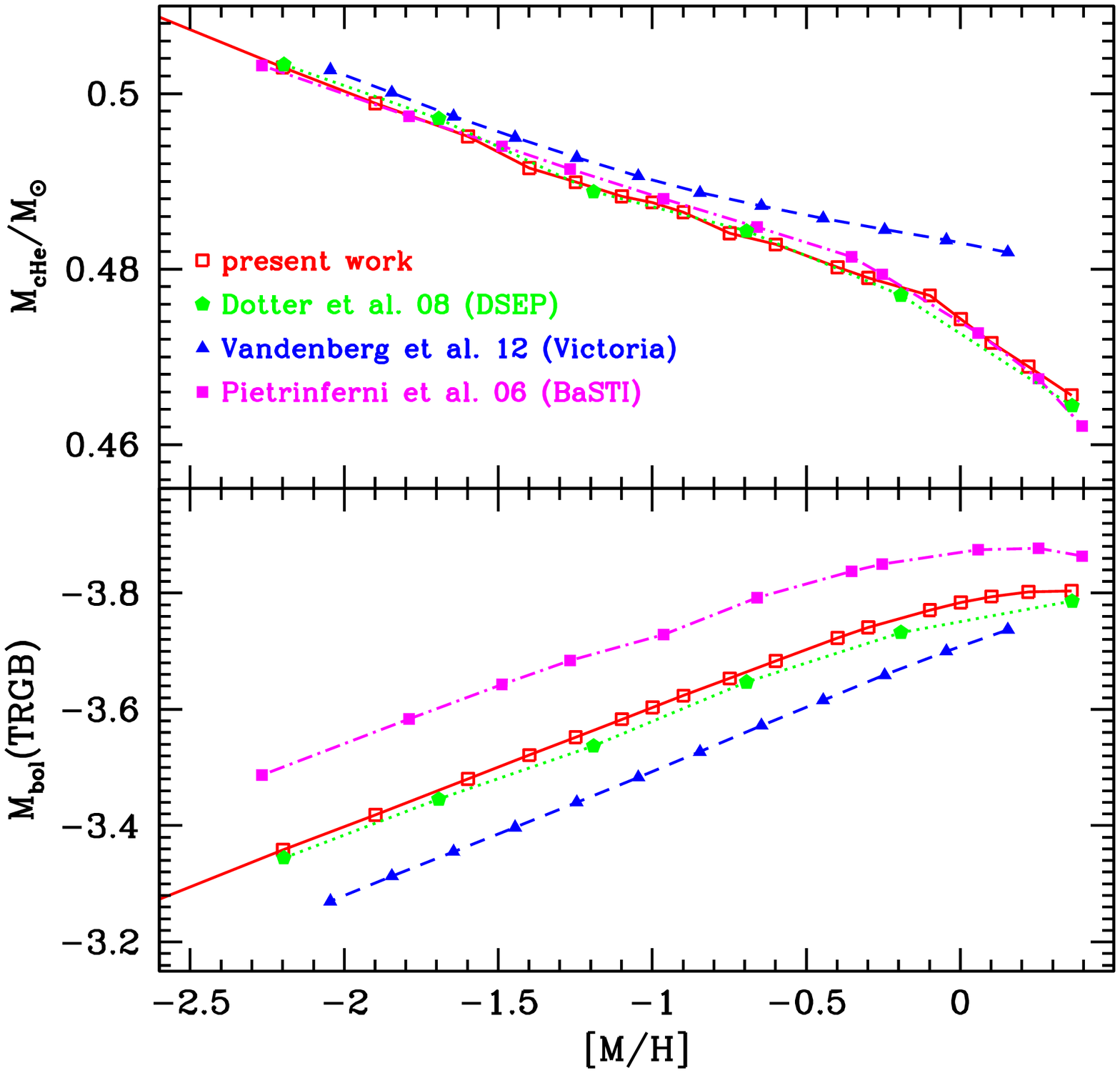}
%\vskip -3.5truecm
\caption{\emph{Upper panel:} TRGB He-core mass (${M_{cHe}}$) as a function of the total metallicity [M/H] for an age of $\sim12.5$~Gyr,
  taken from the labelled  model libraries. \emph{Lower panel:} As the upper panel but for the bolometric luminosity of the TRGB.
 The value $M_{Bol,\odot}=4.74$ of the absolute bolometric magnitude of the Sun has been adopted for all model libraries.\label{fig:tip}}
\end{center}
\end{figure}

\startlongtable
\begin{deluxetable*}{cccccccccc}
  \tablecaption{Initial mass $M_{ini}$, actual mass $M_{fin}$, bolometric luminosity, effective temperature, He-core mass, and 
    absolute magnitudes in the $I$, $J$, $H$, $K$ filters as a function of [Fe/H], for the TRGB of our 12.5~Gyr
    isochrones. \label{tab:tip}}
\tablehead{
\colhead{[Fe/H]} & \colhead{$M_{ini}(M_\odot)$} & \colhead{$M_{fin}(M_\odot)$} & \colhead{$\log(L_{\mathrm TRGB}/L_\odot)$} & \colhead{$\log{T_{\mathrm eff}}$} & \colhead{$M_{cHe}(M_\odot)$} &  \colhead{$M_I$} &  \colhead{$M_J$} &  \colhead{$M_H$} &  \colhead{$M_K$}}
\startdata
  $  -3.20 $  &0.801  &  0.715  &  3.180 & 3.651 &  0.5130  &     $ -3.88$   &$  -4.69$   & $ -5.32$  & $  -5.42$ \\
  $  -2.50 $  &0.800  &  0.706  &  3.239 & 3.646 &  0.5030  & 	$ -4.02$   &$  -4.85$   & $ -5.51$  & $  -5.60$ \\
  $  -2.20 $  &0.801  &  0.701  &  3.264 & 3.637 &  0.4989  & 	$ -4.07$   &$  -4.93$   & $ -5.62$  & $  -5.72$ \\
  $  -1.90 $  &0.806  &  0.699  &  3.288 & 3.623 &  0.4951  & 	$ -4.09$   &$  -5.01$   & $ -5.77$  & $  -5.87$ \\
  $  -1.70 $  &0.812  &  0.699  &  3.305 & 3.612 &  0.4915  & 	$ -4.09$   &$  -5.07$   & $ -5.87$  & $  -5.98$ \\
  $  -1.55 $  &0.818  &  0.701  &  3.317 & 3.603 &  0.4899  & 	$ -4.10$   &$  -5.11$   & $ -5.96$  & $  -6.07$ \\
  $  -1.40 $  &0.826  &  0.703  &  3.329 & 3.592 &  0.4883  & 	$ -4.08$   &$  -5.15$   & $ -6.04$  & $  -6.17$ \\
  $  -1.30 $  &0.833  &  0.707  &  3.337 & 3.585 &  0.4876  & 	$ -4.07$   &$  -5.18$   & $ -6.10$  & $  -6.24$ \\
  $  -1.20 $  &0.840  &  0.711  &  3.345 & 3.577 &  0.4865  & 	$ -4.06$   &$  -5.21$   & $ -6.16$  & $  -6.31$ \\
  $  -1.05 $  &0.855  &  0.720  &  3.357 & 3.566 &  0.4841  & 	$ -4.04$   &$  -5.26$   & $ -6.24$  & $  -6.41$ \\
  $  -0.90 $  &0.874  &  0.735  &  3.369 & 3.555 &  0.4828  & 	$ -4.01$   &$  -5.31$   & $ -6.32$  & $  -6.50$ \\
  $  -0.70 $  &0.905  &  0.759  &  3.385 & 3.537 &  0.4802  & 	$ -3.95$   &$  -5.39$   & $ -6.44$  & $  -6.65$ \\
  $  -0.60 $  &0.923  &  0.775  &  3.392 & 3.528 &  0.4790  & 	$ -3.91$   &$  -5.43$   & $ -6.50$  & $  -6.72$ \\
  $  -0.40 $  &0.967  &  0.818  &  3.404 & 3.511 &  0.4770  & 	$ -3.81$   &$  -5.52$   & $ -6.61$  & $  -6.86$ \\
  $  -0.30 $  &0.981  &  0.830  &  3.410 & 3.501 &  0.4743  & 	$ -3.75$   &$  -5.57$   & $ -6.67$  & $  -6.93$ \\
  $  -0.20 $  &1.003  &  0.853  &  3.414 & 3.492 &  0.4716  & 	$ -3.68$   &$  -5.61$   & $ -6.72$  & $  -7.00$ \\
  $  -0.08 $  &1.027  &  0.879  &  3.417 & 3.481 &  0.4689  & 	$ -3.59$   &$  -5.67$   & $ -6.78$  & $  -7.07$ \\
  $  +0.06 $  &1.054  &  0.908  &  3.418 & 3.468 &  0.4656  &	$ -3.46$   &$  -5.72$   & $ -6.83$  & $  -7.15$ \\
\enddata
\end{deluxetable*}

Figure~\ref{fig:tip} shows bolometric magnitude and He-core mass at the TRGB (${M_{cHe}}$) of our new isochrones, compared with
the values from \cite{basti:06}, DSEP, and Victoria \citep{don:12} models at a reference age
of 12.5~Gyr, as a function of the total metallicity [M/H]. We use [M/H] in this comparison to
minimize the effect of 
different reference solar heavy element distributions and [$\alpha$/Fe]\footnote{Victoria models are
  calculated with the \citet{gs:98} solar metal distribution, and an $\alpha$-enhancement varying from
element to element, in the range between 0.25 and 0.5~dex.}.
%(Yonsei-Yale models are calculated for [$\alpha$/Fe]=+0.3, instead of [$\alpha$/Fe]=+0.4).
There are however several differences in physics inputs and initial $Y$ among these sets of models. 

The values of $M_{cHe}$ in our new models and \cite{basti:06} results are very similar, with a difference of only
about 0.002~$M_\odot$ between [M/H]$\sim -$1.6 and [M/H]$\sim -$0.3, the previous BaSTI values being higher.
In general, the models compared in Fig.~\ref{fig:tip} display $M_{cHe}$ values within about $\pm$0.004~$M_\odot$ around
our results, the exception being Victoria models,
which for [M/H]$> -$0.9 provide increasingly higher values compared to ours.
This latter behaviour is likely due to the assumption of constant $Y$ at all metallicities in the Victoria models,
while all other calculations have $Y$ increasing with $Z$. An increase of $Y$ at fixed $Z$ tends to decrease ${M_{cHe}}$, and
this explains at least qualitatively the comparison with Victoria models
\footnote{It needs also to be noted that in the Victoria
  calculations, ${M_{cHe}}$ is defined as the mass enclosed  between the centre and the mid-point of the H-burning shell,
  whilst in the other models ${M_{cHe}}$ is
  taken as the mass size of the region where H has been exhausted. This different definition can contribute
  to explain the residual difference in the low metallicity
  regime \citep[see,][for a detailed discussion on this issue]{don:12}.}.

\begin{figure}[ht!]
\begin{center}
\includegraphics[width=3.4in]{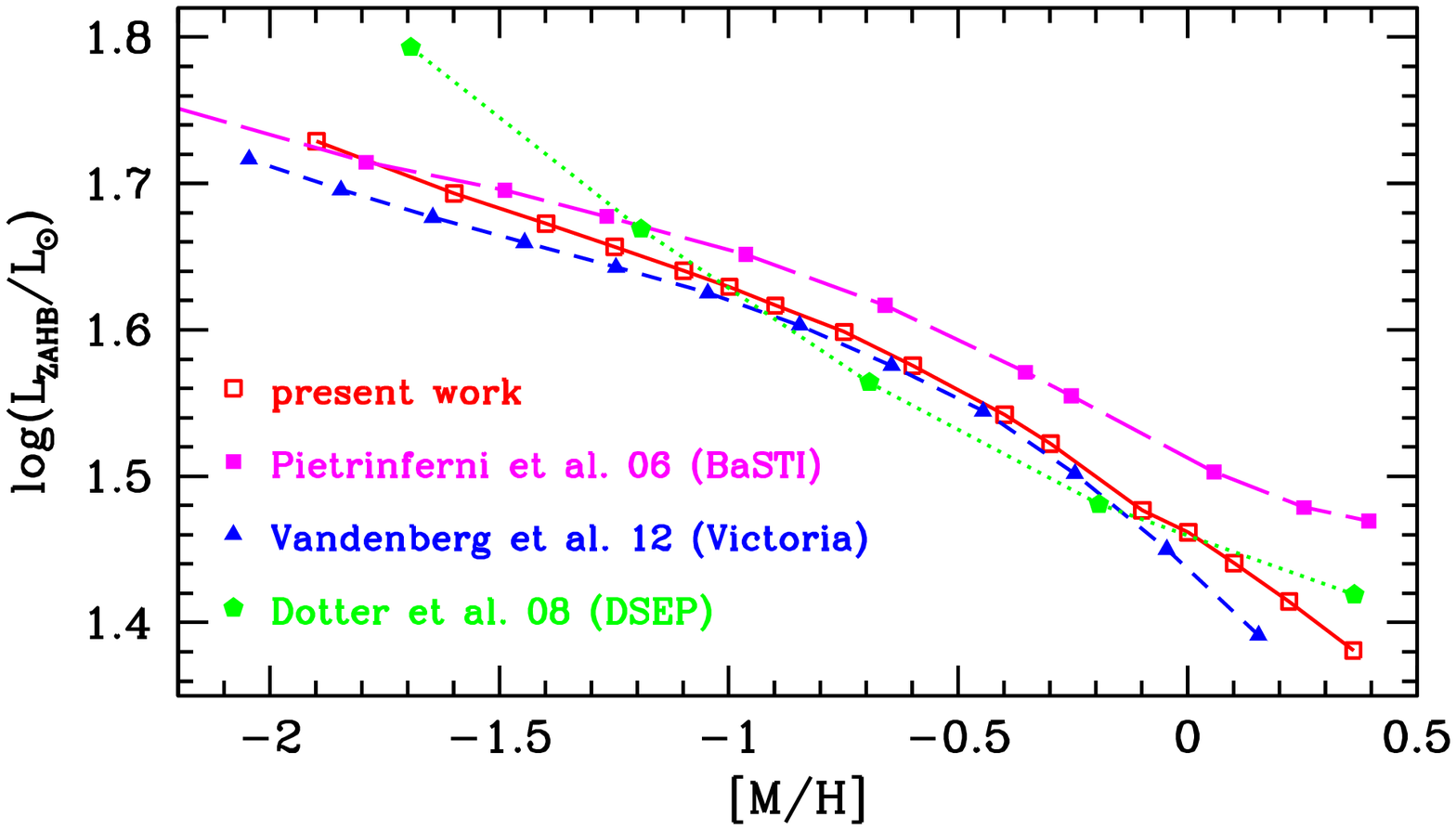}
\vskip -2.5truecm
\caption{The luminosity of ZAHB models at the RR Lyrae instability strip (taken at $\log{T_{\mathrm eff}}=3.83$),
  as a function of the total
  metallicity [M/H], from the labelled model libraries.
\label{fig:hbcomp}}
\end{center}
\end{figure}

Regarding the TRGB, Fig.~\ref{fig:tip} shows that our new models are significantly fainter -- by about $\sim0.15$~mag at
[M/H]=$-1.3$ -- than the previous BaSTI predictions. This is partially due to the
smaller He-core mass in the new calculations, but the main reason is the different reference solar metal distribution,
and the inclusion of atomic diffusion. DSEP models are only slightly
underluminous, and 
%the Yonsei-Yale TRGB luminosities are slightly fainter than DSEP ones apart from the highest metallicities, at
%which they are substantially fainter than our results.
Victoria models are consistently fainter by about 0.1~mag, despite their larger ${M_{cHe}}$. This is partially due
the different definition  concerning $M_{cHe}$ as well as to the lower initial He abundance adopted in the Victoria model library.

As a reference, Table~\ref{tab:tip} reports several quantities at the TRGB as a function of [Fe/H] from our calculations,
including ${M_{cHe}}$,
the bolometric luminosity of the TRGB and absolute magnitudes in selected filters used for the TRGB distance scale
(see next section).

Figure~\ref{fig:hbcomp} compares the luminosity of our new
ZAHB models at the RR Lyrae instability strip
(taken at $\log{T_{\mathrm eff}}=3.83$) with \citep{basti:06}, DSEP and Victoria results, again as a function of the total
metallicity [M/H].
Victoria models give results very close to ours (only slightly underluminous), whilst luminosities
from the older BaSTI models are typically higher, with differences increasing for increasing [M/H].
DSEP models are much brighter at the lowest metallicities, then become close to our calculations. On the whole,
at [M/H] between $\sim -$1.2 and 0.0, DSEP and Victoria models are
consistent with our new results within about $\pm$0.02 dex.
Table~\ref{tab:zahb} summarizes some main properties of our ZAHB models at the RR Lyrae instability strip.

\startlongtable
\begin{deluxetable*}{cccccccccc}
  \tablecaption{Bolometric luminosity, absolute magnitudes in the $BVRIJHK$ filters, and the mass $M_{RR}$ of the ZAHB model at 
  $\log{T_{\mathrm eff}}=3.83$ --taken as representative of the average $T_{\mathrm eff}$ within the RR Lyrae instability strip \citep[][]{marconi15}-- as a function  of [Fe/H]. \label{tab:zahb}}
\tablehead{
\colhead{[Fe/H]} & \colhead{$M_{RR}(M_\odot)$}  & \colhead{$\log(L_{\mathrm ZAHB}/L_\odot)$} &  \colhead{$M_B$} & \colhead{$M_V$} &   \colhead{$M_R$} & \colhead{$M_I$} &  \colhead{$M_J$} &  \colhead{$M_H$} &  \colhead{$M_K$}     }
\startdata
   $-2.20$  &  0.791 & 1.728  &   0.709 &  0.42  & 0.20 & $  -0.03$ & $-0.31$ &  $-0.52$ & $-0.54 $\\
   $-1.90$  &  0.723 & 1.691  &   0.794 &  0.50  & 0.28 & $ ~0.06$ & $-0.22$ &  $-0.43$ & $-0.45  $\\
   $-1.70$  &  0.687 & 1.670  &   0.847 &  0.55  & 0.33 & $ ~0.11$ & $-0.17$ &  $-0.38$ & $-0.40  $\\
   $-1.55$  &  0.665 & 1.654  &   0.886 &  0.58  & 0.37 & $ ~0.15$ & $-0.13$ &  $-0.34$ & $-0.37  $\\
   $-1.40$  &  0.646 & 1.641  &   0.915 &  0.61  & 0.39 & $ ~0.17$ & $-0.10$ &  $-0.31$ & $-0.33  $\\
   $-1.30$  &  0.635 & 1.627  &   0.950 &  0.64  & 0.43 & $ ~0.21$ & $-0.07$ &  $-0.27$ & $-0.30  $\\
   $-1.20$  &  0.624 & 1.615  &   0.979 &  0.67  & 0.45 & $ ~0.23$ & $-0.04$ &  $-0.24$ & $-0.27  $\\
   $-1.05$  &  0.608 & 1.596  &   1.023 &  0.71  & 0.49 & $ ~0.27$ & $ ~0.00$ &  $-0.20$ & $-0.23 $\\
   $-0.90$  &  0.596 & 1.574  &   1.086 &  0.76  & 0.54 & $ ~0.33$ & $ ~0.06 $ & $-0.15 $ &$-0.17 $\\
   $-0.70$  &  0.580 & 1.541  &   1.155 &  0.83  & 0.61 & $ ~0.39$ & $ ~0.13$ &  $-0.07$ & $-0.09 $\\
   $-0.60$  &  0.573 & 1.520  &   1.227 &  0.88  & 0.66 & $ ~0.45$ & $ ~0.18$ &  $-0.01$ & $-0.04 $\\
   $-0.40$  &  0.562 & 1.475  &   1.334 &  0.98  & 0.75 & $ ~0.55$ & $ ~0.28$ &  $ ~0.09 $ & $~0.06 $\\
   $-0.30$  &  0.555 & 1.458  &   1.404 &  1.03  & 0.81 & $ ~0.60$ & $ ~0.33$ &  $ ~0.14 $ & $~0.11 $\\
   $-0.20$  &  0.547 & 1.439  &   1.424 &  1.06  & 0.83 & $ ~0.62$ & $ ~0.36$ & $ ~0.18 $  &$~0.15 $\\
   $-0.08$  &  0.539 & 1.413  &   1.483 &  1.11  & 0.88 & $ ~0.67$ & $ ~0.42$ & $ ~0.24 $  &$~0.21 $\\
   $+0.06$  &  0.531 &  1.381 &   1.571 &  1.18  & 0.95 & $ ~0.75 $ & $~0.49 $ & $~0.31 $  &$~0.29 $\\
\enddata
\end{deluxetable*}

\section{Comparisons with observations}
\label{obscomp}

In this section, we discuss results of a few tests to assess the general consistency of our new models
and isochrones with selected observational constraints.
In all of these comparisons, we have included the effect of extinction according to the standard
\citet{ccm89} reddening law, with $R_V\equiv A_V/E(B-V)$ = 3.1, and have calculated the ratios $A_{\lambda}/A_V$
for the relevant photometric filters, as described in \citet{Girardi02}\footnote{For the comparison with the 
  {\em HST/ACS} photometry of NGC~6397
  we have taken into account the dependence of the extinction ratios on $T_{\mathrm eff}$,
  because it is much stronger than in Johnson-Cousins filters.}

\begin{figure}[ht!]
\begin{center}
\includegraphics[width=3.4in]{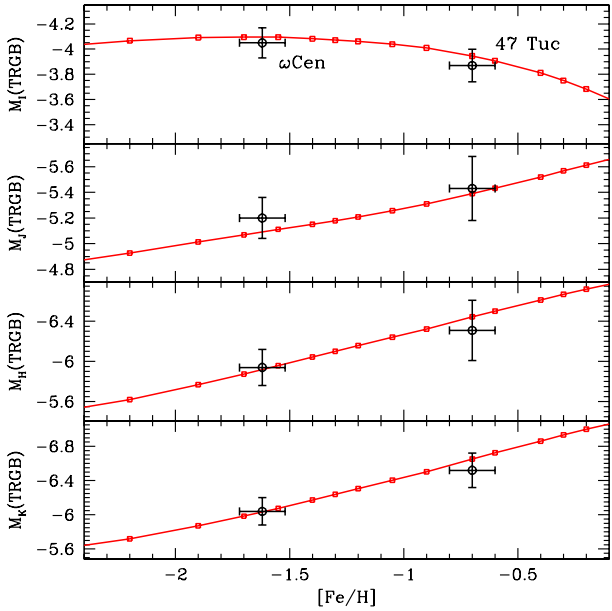}
\caption{Comparison of the TRGB absolute magnitudes of our new calculations (see Table~\ref{tab:tip}) with
  the $IJHK$ empirical results by \citet{bell04}
  for the Galactic globular clusters 47~Tuc and $\omega$~Centauri (see text for details).\label{fig:tipGC}}
\end{center}
\end{figure}

The first test is shown in Fig.~\ref{fig:tipGC}, which displays a comparison between the new 
theoretical $IJHK$ TRGB absolute magnitudes of Table~\ref{tab:tip}, and the empirical results
for 47~Tuc and $\omega$~Centauri by \citet{bell04}. Their derived absolute magnitudes for the
TRGB of 47~Tuc have been shifted by +0.04~mag, to account for the new eclipsing binary distance 
by \citet{eb20}. The values for $\omega$~Centauri are unchanged, because already based on
an eclipsing binary distance to this cluster \citep[see discussion in][]{bell04}. The metallicity
assigned to $\omega$~Centauri is the [Fe/H] of the main cluster population, as discussed in \citet{bell04}.
Our theoretical TRGB magnitudes appear nicely consistent with these results in all filters,
within the corresponding error bars.

The following Fig.\ref{fig:zahbGC} displays a comparison of the model
ZAHB luminosities reported in Table~\ref{tab:zahb} with
the semiempirical results by \citet{dc:99}, based on the pulsational properties of RR Lyrae stars in
GCs. Also in this case we find a general consistency with our models\footnote{An important ingredient
entering the analysis by \citet{dc:99} is the range of masses that populate the RR Lyrae instability strip. 
Given that this quantity was determined using stellar models, we have verified that our new calculations
do not change the mass ranges employed by \citet{dc:99}.}

\begin{figure}[ht!]
\begin{center}
\includegraphics[width=3.5in]{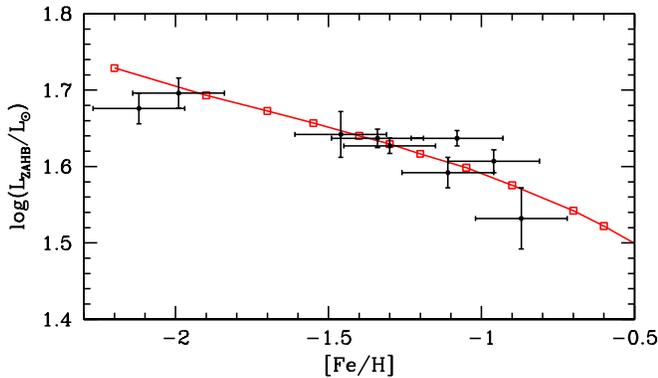}
\vskip -3truecm
\caption{As Fig.~\ref{fig:hbcomp}, but showing our new calculations compared to the
  semiempirical results by \citet{dc:99}.  \label{fig:zahbGC}}
\end{center}
\end{figure}

\begin{figure}[ht!]
\begin{center}
\includegraphics[width=3.5in]{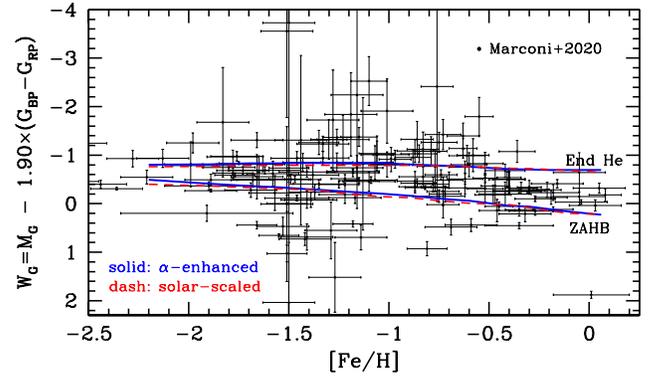}
\vskip -3truecm
\caption{Wesenheit $W_G$ - [Fe/H] diagram of field RR Lyrae stars
  from {\em Gaia} data Release 2 \citet{marconi20} compared to the 
  ZAHB and the sequence corresponding to the exhaustion of He in the center, 
  for our $\alpha$-enhanced (solid lines) and solar scaled models (dashed lines -- see text for details).
  \label{fig:rrgaia}}
\end{center}
\end{figure}

We have also compared our HB models with {\em Gaia} Data Release 2 results for 
  a sample of Galactic field RR Lyrae stars with accurate parallaxes, magnitudes, and
  high-resolution spectroscopic measurements of [Fe/H],
  taken from \cite{marconi20} (their table A1).
  To remove uncertainties associated with both the poorly constrained extinction and
  $\alpha-$element enhancement, Fig.~\ref{fig:rrgaia} shows
  the relation between the measured values of the reddening-free Wesenheit
  index $W_{G} = M_{G} - 1.90\times(G_{BP}-G_{RP})$ \citep[see][]{ripepi19} and the iron abundance of the
  stars in the sample.
  This data is compared to the corresponding relationship - taken at  $\log{T_{\mathrm eff}}=3.83$
  (see previous discussion) - predicted by both $\alpha$-enhanced and solar
  scaled \citep[from][]{bastiac:18}  ZAHB models, that display almost identical $W_{G}$ values
  at fixed [Fe/H].
  The evolution off-ZAHB of the tracks
  in the instability strip display a small increase of $W_{G}$ by at most $\sim$0.2~mag at intermediate and high
  [Fe/H], followed by a steady decrease until the exhaustion of central He (the sequences corresponding to 
  the exhaustion of central He are also displayed). 
  But for a few peculiar cases that would need to be analyzed individually,
  the large majority of the stars across the whole [Fe/H] range
  lie either slightly below (fainter $W_{G}$) the ZAHB --consistent with the
  evolutionary path of the tracks-- or above it, as expected from the models. 
  There are also several objects located above the sequences corresponding to the exhaustion of central He, but 
    we refrain from speculating about their origin, given the large errors on $W_{G}$ that affect most of these stars.
  
\begin{figure}[ht!]
\begin{center}
\includegraphics[width=3.4in]{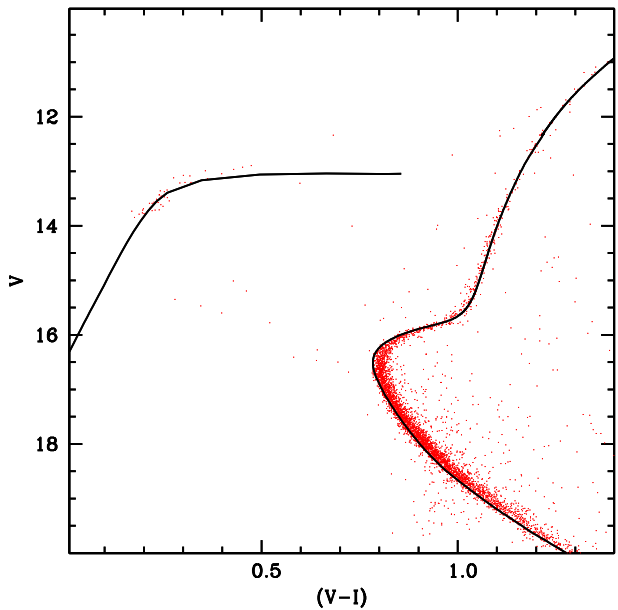}
\caption{Comparison of a [Fe/H]=$-$1.9, 13.5~Gyr isochrone and ZAHB with $Y$=0.248 (solid line)  
  with the \citet{st00} $VI$ CMD of NGC~6397 (see text for details).\label{fig:fit6397j}}
\end{center}
\end{figure}

\begin{figure}[ht!]
\begin{center}
\includegraphics[width=3.4in]{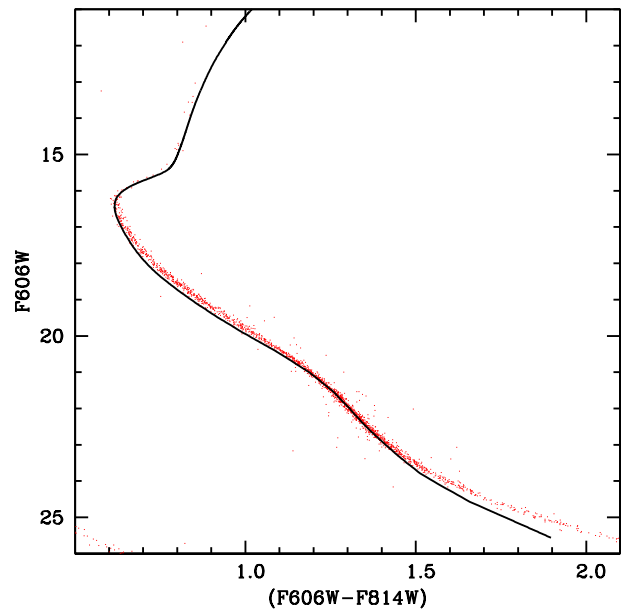}
\caption{Comparison of a [Fe/H]=$-$1.9, 13.5~Gyr isochrone with $Y$=0.248 (solid line)  
  with the {\em HST/ACS} CMD of NGC~6397 by \citet{richer08} (see text for details).\label{fig:fit6397acs}}
\end{center}
\end{figure}

A test with CMDs of the metal poor Galactic globular clusters (GGC) NGC~6397 ([Fe/H]=$-2.03 \pm 0.05$,[$\alpha$/Fe]=0.34$\pm$0.02) \citep[see][]{gr03} follows in Figs.~\ref{fig:fit6397j} and ~\ref{fig:fit6397acs}.
The first comparison is between our [Fe/H]=$-$1.9, $Y$=0.248 isochrones and ZAHB models, and the 
Johnson-Cousins $VI$ photometry by \citet{st00}, as shown in Fig.~\ref{fig:fit6397j}. 
The fit of the theoretical ZAHB and the lower MS to the observed CMD constrain distance modulus and reddening
to $E(B-V)$=0.19 and $(m-M)_0$=11.96, respectively.
The TO region is matched by a 13.5~Gyr isochrone, that is also nicely consistent with
the observed RGB.
These values of reddening and distance well agree with 
$E(B-V)=0.183 \pm 0.005 ({\rm stat}) \pm 0.011 ({\rm syst})$ estimated by \citet{gr03},
and $(m-M)_0 = 11.89 \pm 0.07 ({\rm stat}) \pm 0.09 ({\rm syst})$ determined by 
\citet{br18}, from measurements of the cluster parallax distance using
the {\em HST/WFC3} spatial-scanning mode.

Like most GGCs, stars in this cluster displays the well known O-Na and C-N abundance anticorrelations
\citep[see, e.g.,][for a review on the topic]{bl18}, usually associated also
to a range of helium abundances. The anticorrelations do not affect isochrones and bolometric corrections in optical filters,
but the initial helium abundance does, through its effect on model luminosities, lifetimes and $T_{\mathrm eff}$  
 \citep[see, e.g.,][for a review]{cs20}. However, the He abundance spread is negligible in this cluster  
 \citep[as derived by][]{mi18} and isochrones for a single, standard value of $Y$ are appropriate to match
 the observed CMD.

The deep {\em HST/ACS} optical CMD from \citet{richer08} is displayed in Fig.~\ref{fig:fit6397acs}, together with
the same isochrone of Fig.~\ref{fig:fit6397j} but in the filter system of the ACS camera on board of {\em HST},
using the same distance moduls and reddening
of the comparison in the Johnson-Cousins CMD.
In this case, we note that between $\sim$1 and $\sim$5 $F606W$ magnitudes below the TO the
  isochrone is systematically bluer than the data; the same happens when $F606W$ is fainter than  
  $\sim$7 magnitudes below the TO. While this latter discrepancy is found also for the higher metallicity example discussed
  below (see the discussion on 47~Tuc that will follow) the same is not true for the brighter magnitude range.
  The reason might be related to possible metallicity-dependent offsets of the bolometric corrections for the
  {\em HST/ACS} system, but comparisons with more clusters are required to reach a definitive conclusion.
%The extended MS is reasonably well matched,
%at least until $F606W\sim 23.5$, .

\begin{figure}[ht!]
\begin{center}
\includegraphics[width=3.4in]{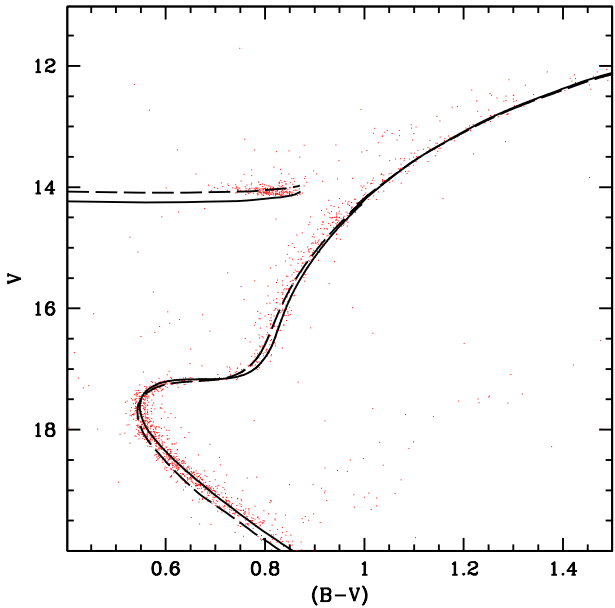}
\caption{Fit of two [Fe/H]=$-$0.7, 12.3~Gyr isochrones and ZAHBs with $Y$=0.255 (solid line)
  and 0.300 (dashed line), respectively, 
  to the \citet{bs09} $BV$ CMD of 47~Tuc  (see text for details).\label{fig:fit47Tuc}}
\end{center}
\end{figure}

The next comparison is with the $BV$ CMD \citep[from][]{bs09}
of the metal rich GGC 47~Tuc ([Fe/H]=$-0.66 \pm 0.04$, [$\alpha$/Fe]=0.30$\pm$0.02) \citep[see][]{gr03}.
  This cluster has an internal He abundance spread with a range
  $\Delta Y\sim$0.03-0.05 \citep{dicr10,scp16,mi18}, and we use ZAHB and isochrones for both normal $Y$=0.255
  and enhanced $Y$=0.300 helium.
  We fix simultaneously the distance modulus $(m-M)_{0}=13.30$ and reddening $E(B-V)$=0.02, 
  by matching the lower envelope of the red HB
  and approximately the red edge of the lower MS, with ZAHB and isochrones for $Y$=0.255.
  We then tested that, for the same reddening and distance,
  the helium enhanced isochrones and ZAHB are still consistent with the observed sequences
  in the CMD.
  Figure~\ref{fig:fit47Tuc} displays a comparison between the cluster 
  CMD and 12.3~Gyr, [Fe/H]=$-$0.7, $Y$=0.255 and $Y$=0.300 isochrones, which
  match the position of the cluster TO, together with ZAHB models for both helium abundances.
  The derived value of $E(B-V)$ is in excellent agreement with $E(B-V)=0.024 \pm 0.004 ({\rm stat}) \pm 0.011 ({\rm syst})$ estimated by \citet{gr03}; the distance  is fully consistent with the average $(m-M)_{0}=13.27 \pm 0.07$ obtained from
  two cluster eclipsing binaries \citep{eb10, eb20}.

Another empirical and independent distance determination for this cluster, based on 
{\em Gaia} Data Release 2 results, provides
$(m-M)_{0}=13.24\pm 0.005 ({\rm stat})\pm0.058 ({\rm sys})$~mag ~\citep{chen18}, consistent with both our result and
the eclipsing binary analysis.

\begin{figure}[ht!]
\begin{center}
\includegraphics[width=3.4in]{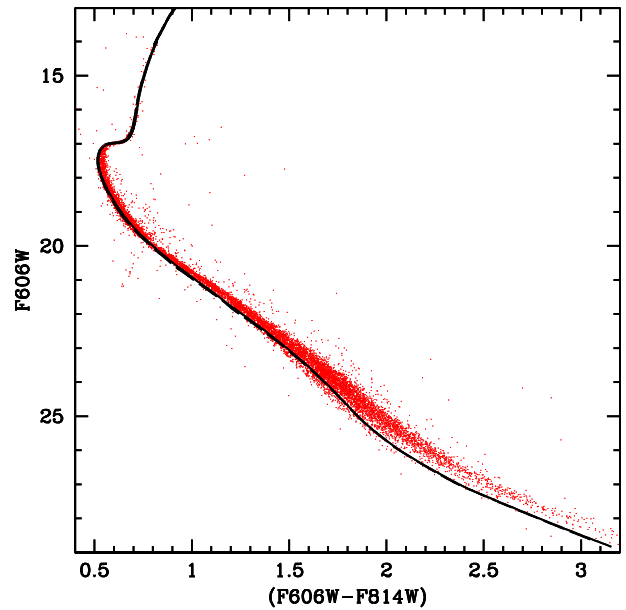}
\caption{As Fig.~\ref{fig:fit47Tuc} but for \citet{kal}
  HST/ACS CMD (see text for details).\label{fig:fit47Tucvlm}}
\end{center}
\end{figure}

\begin{figure}[ht!]
\begin{center}
\includegraphics[width=3.4in]{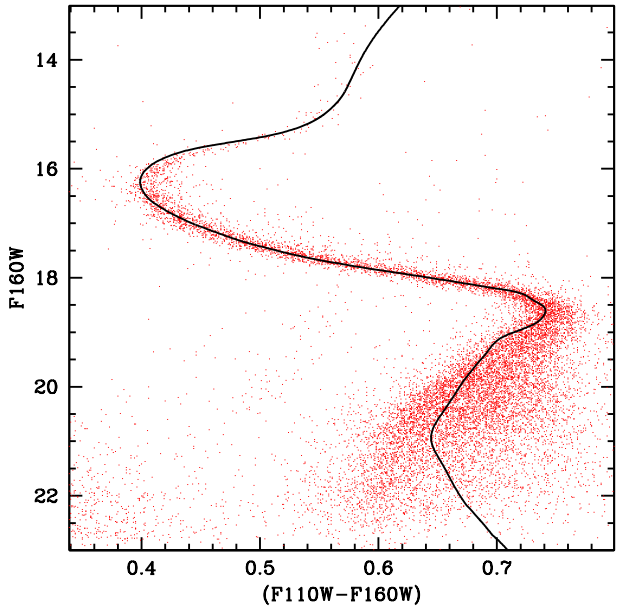}
\caption{As Fig.~\ref{fig:fit47Tuc} but for
  the {\em HST/WFC3} infrared CMD by \citet{kal}. Only the $Y$=0.255 isochrone is shown 
  (see text for details).\label{fig:fit47Tucir}}
\end{center}
\end{figure}

Figure~\ref{fig:fit47Tucvlm} displays the much deeper {\em HST/ACS} optical CMD
by \citet{kal} compared to the same isochrones of Fig.~\ref{fig:fit47Tuc}, using the same distance modulus
and extinction. As for the case of the more metal poor cluster NGC~6397, the fainter part of the
isochrone  MS (for both values of the initial helium)
is systematically bluer than the observations. To investigate this issue, we show in 
Fig.~\ref{fig:fit47Tucir} the {\em HST/WFC3} infrared CMD by \citet{kal}, compared 
only to the $Y$=0.255 isochrone, using again the same distance modulus and extinction of Fig.~\ref{fig:fit47Tuc}. 
In this CMD the isochrone does not appear systematically bluer than the data along the lower MS, and
follows well the observed changing shape, due to the competition between
the collision induced absorption of the ${\rm H_2}$ molecule in the infrared (that shifts
the colours to the blue) and the increase of the radiative opacity with decreasing $T_{\mathrm eff}$
\citep[see, e.g.,][and references therein]{cs:13}. This suggests that the systematic difference between
theory and observations
found in optical colours might be due to the adopted bolometric corrections.
Below $F160W \sim$18.5 the ($F110W-F160W$) colour is sensitive to the specific
metal abundance patterns of the He-enhanced multiple
populations hosted by the cluster, that affect the bolometric corrections.
As shown by \citet{mil}, the result is to have redder colours at fixed
$F160W$, compared to models with standard $\alpha$-enhanced composition.

\begin{figure}[ht!]
\begin{center}
\includegraphics[width=3.4in]{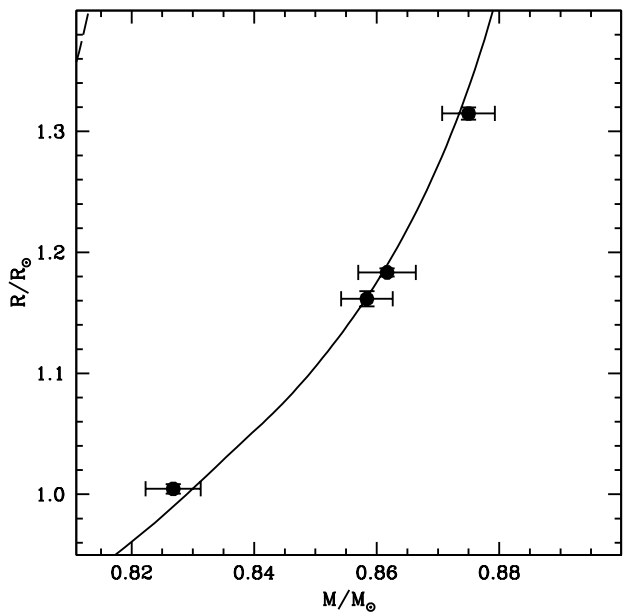}
\caption{Comparison of an isochrone for a [Fe/H]=$-$0.7, $Y$=0.255, 12.3~Gyr, with data for the cmponents of two eclipsing binaries
  in 47~Tuc, in a mass-radius diagram (see text for details).\label{fig:fit47TucEB}}
\end{center}
\end{figure}

Figure~\ref{fig:fit47TucEB} shows the 12.3~Gyr $Y$=0.255 isochrone in a mass-radius diagram, compared to
the masses and radii of the components of the two cluster eclipsing binaries. The age determined from the CMD is
nicely consistent with the radius of the eclipsing binary components.
The isochrone for $Y$=0.300 lies outside the boundaries of this diagram, shifted to
masses too low to be consistent with the data.

\section{Conclusions}
\label{conclusions}

We have presented an overview of the updated BaSTI $\alpha$-enhanced models, whose input physics and reference
solar metal mixture are consistent with the updated solar scaled models of Paper~I.
Like for the new solar scaled models, the updated $\alpha$-enhanced library increases
significantly the number of available metallicities, includes the very low-mass star regime,
accounts consistently for the pre-MS evolution
in the isochrone calculations, and also provides the asteroseismic properties of the models.

We successfully
tested these new calculations against the luminosities of the ZAHB and TRGB
in selected GGCs. We also compared the isochrones with 
CMDs of one metal rich (47~Tuc) and one metal poor (NGC~6397) GGC; they provide a good fit to the observed CMDs,
for distance moduli consistent with both the parallax and eclipsing binary distance to  47~Tuc,
and the parallax distance to NGC~6397.
The best fit isochrone for 47~Tuc can also nicely match the mass-radius diagram
of the components of two cluster eclipsing binaries. 

Like for the updated solar scaled library,
the entire $\alpha$-enhanced database is publicly available at the following dedicated websites:
{\url http://basti-iac.oa-abruzzo.inaf.it} and {\url https://basti-iac.iac.es}.
Here we include stellar
evolution tracks and isochrones in several photometric systems, and 
the asteroseismic properties of our grid of stellar evolution calculations.
We can also provide, upon request, additional calculations (both evolutionary and
asteroseismic
outputs) for masses not included in our standard grids. These websites include also a web-tool to
calculate online synthetic CMDs for any arbitrary star formation history and age metallicity relation, using
the updated BaSTI isochrones. Details about the inputs to specify when running this
web-tool, as well as a detailed discussion of the outputs, are provided in the Appendix.

\acknowledgements{We warmly thank the referee for the very constructive comments that have greatly improved 
    the presentation of our results.
    We also wish to warmly thank M. Correnti, J. Kalirai, M. Marconi, and V. Ripepi for sharing the results
    of their research. AP and SC acknowledge financial support from \lq{Progetto Premiale}\rq\ MIUR {\sl MITIC} (PI: B. Garilli)}, and progetto INAF Mainstream (PI: S. Cassisi). SC acknowledges support from {\em PLATO} ASI-INAF agreement n.2015-019-R.1-2018, and from INFN (Iniziativa specifica TAsP). This research has been supported by the Spanish Ministry of Science and Innovation (P.I. S. Hidalgo) under the grant AYA2017-89841-P.

\newpage
\bibliographystyle{aasjournal.bst}
\bibliography{basti_aep04}

\begin{thebibliography}{}
\expandafter\ifx\csname natexlab\endcsname\relax\def\natexlab#1{#1}\fi

\bibitem[{{Allard} {et~al.}(2012){Allard}, {Homeier}, \& {Freytag}}]{allard:12}
{Allard}, F., {Homeier}, D., \& {Freytag}, B. 2012, Philosophical Transactions
  of the Royal Society of London Series A, 370, 2765

\bibitem[{{Bastian} \& {Lardo}(2018)}]{bl18}
{Bastian}, N., \& {Lardo}, C. 2018, \araa, 56, 83

\bibitem[{{Bellazzini} {et~al.}(2004){Bellazzini}, {Ferraro}, {Sollima},
  {Pancino}, \& {Origlia}}]{bell04}
{Bellazzini}, M., {Ferraro}, F.~R., {Sollima}, A., {Pancino}, E., \& {Origlia},
  L. 2004, \aap, 424, 199

\bibitem[{{Bergbusch} \& {Stetson}(2009)}]{bs09}
{Bergbusch}, P.~A., \& {Stetson}, P.~B. 2009, \aj, 138, 1455

\bibitem[{{Bergemann} \& {Serenelli}(2014)}]{bergemann:14}
{Bergemann}, M., \& {Serenelli}, A. 2014, {Solar Abundance Problem}, ed.
  E.~{Niemczura}, B.~{Smalley}, \& W.~{Pych}, 245--258

\bibitem[{{Bessell} \& {Murphy}(2012)}]{Bessell12}
{Bessell}, M., \& {Murphy}, S. 2012, \pasp, 124, 140

\bibitem[{{Bessell}(1990)}]{Bessel90}
{Bessell}, M.~S. 1990, \pasp, 102, 1181

\bibitem[{{Bessell}(2011)}]{Bessell11}
---. 2011, Astronomical Society of the Pacific Conference Series, Vol. 451,
  {Science with the Skymapper Telescope}, ed. S.~{Qain}, K.~{Leung}, L.~{Zhu},
  \& S.~{Kwok}, 323

\bibitem[{{Bessell} \& {Brett}(1988)}]{Bessel88}
{Bessell}, M.~S., \& {Brett}, J.~M. 1988, \pasp, 100, 1134

\bibitem[{{Bessell} {et~al.}(1998){Bessell}, {Castelli}, \& {Plez}}]{Bessel98}
{Bessell}, M.~S., {Castelli}, F., \& {Plez}, B. 1998, \aap, 333, 231

\bibitem[{{Brown} {et~al.}(2018){Brown}, {Casertano}, {Strader}, {Riess},
  {VandenBerg}, {Soderblom}, {Kalirai}, \& {Salinas}}]{br18}
{Brown}, T.~M., {Casertano}, S., {Strader}, J., {et~al.} 2018, \apjl, 856, L6

\bibitem[{{Cabrera-Ziri} {et~al.}(2020){Cabrera-Ziri}, {Speagle},
  {Dalessandro}, {Usher}, {Bastian}, {Salaris}, {Martocchia},
  {Kozhurina-Platais}, {Niederhofer}, {Lardo}, {Larsen}, \&
  {Saracino}}]{cabrera20}
{Cabrera-Ziri}, I., {Speagle}, J.~S., {Dalessandro}, E., {et~al.} 2020, \mnras,
  495, 375

\bibitem[{{Caffau} {et~al.}(2011){Caffau}, {Ludwig}, {Steffen}, {Freytag}, \&
  {Bonifacio}}]{caffau:11}
{Caffau}, E., {Ludwig}, H.-G., {Steffen}, M., {Freytag}, B., \& {Bonifacio}, P.
  2011, \solphys, 268, 255

\bibitem[{{Cardelli} {et~al.}(1989){Cardelli}, {Clayton}, \& {Mathis}}]{ccm89}
{Cardelli}, J.~A., {Clayton}, G.~C., \& {Mathis}, J.~S. 1989, \apj, 345, 245

\bibitem[{{Cassisi} {et~al.}(1998){Cassisi}, {Castellani}, {degl'Innocenti}, \&
  {Weiss}}]{cassisi:98}
{Cassisi}, S., {Castellani}, V., {degl'Innocenti}, S., \& {Weiss}, A. 1998,
  \aaps, 129, 267

\bibitem[{{Cassisi} {et~al.}(2007){Cassisi}, {Potekhin}, {Pietrinferni},
  {Catelan}, \& {Salaris}}]{cassisi:07}
{Cassisi}, S., {Potekhin}, A.~Y., {Pietrinferni}, A., {Catelan}, M., \&
  {Salaris}, M. 2007, \apj, 661, 1094

\bibitem[{{Cassisi} \& {Salaris}(2013)}]{cs:13}
{Cassisi}, S., \& {Salaris}, M. 2013, {Old Stellar Populations: How to Study
  the Fossil Record of Galaxy Formation}

\bibitem[{{Cassisi} \& {Salaris}(2020)}]{cs20}
---. 2020, \aapr, 28, 5

\bibitem[{{Cassisi} {et~al.}(2004){Cassisi}, {Salaris}, {Castelli}, \&
  {Pietrinferni}}]{cassisi:04}
{Cassisi}, S., {Salaris}, M., {Castelli}, F., \& {Pietrinferni}, A. 2004, \apj,
  616, 498

\bibitem[{{Cayrel de Strobel} {et~al.}(1997){Cayrel de Strobel}, {Crifo}, \&
  {Lebreton}}]{cayrel}
{Cayrel de Strobel}, G., {Crifo}, F., \& {Lebreton}, Y. 1997, in ESA Special
  Publication, Vol. 402, Hipparcos - Venice '97, ed. R.~M. {Bonnet},
  E.~{H{\o}g}, P.~L. {Bernacca}, L.~{Emiliani}, A.~{Blaauw}, C.~{Turon},
  J.~{Kovalevsky}, L.~{Lindegren}, H.~{Hassan}, M.~{Bouffard}, B.~{Strim},
  D.~{Heger}, M.~A.~C. {Perryman}, \& L.~{Woltjer}, 687--688

\bibitem[{{Chen} {et~al.}(2018){Chen}, {Richer}, {Caiazzo}, \& {Heyl}}]{chen18}
{Chen}, S., {Richer}, H., {Caiazzo}, I., \& {Heyl}, J. 2018, \apj, 867, 132

\bibitem[{{Christensen-Dalsgaard}(2008)}]{oscicode}
{Christensen-Dalsgaard}, J. 2008, \apss, 316, 113

\bibitem[{{Cohen} {et~al.}(2003){Cohen}, {Wheaton}, \& {Megeath}}]{Cohen03}
{Cohen}, M., {Wheaton}, W.~A., \& {Megeath}, S.~T. 2003, \aj, 126, 1090

\bibitem[{{Cordier} {et~al.}(2007){Cordier}, {Pietrinferni}, {Cassisi}, \&
  {Salaris}}]{basti:07}
{Cordier}, D., {Pietrinferni}, A., {Cassisi}, S., \& {Salaris}, M. 2007, \aj,
  133, 468

\bibitem[{{de Santis} \& {Cassisi}(1999)}]{dc:99}
{de Santis}, R., \& {Cassisi}, S. 1999, \mnras, 308, 97

\bibitem[{{De Somma} {et~al.}(2020){De Somma}, {Marconi}, {Molinaro},
  {Cignoni}, {Musella}, \& {Ripepi}}]{desomma20}
{De Somma}, G., {Marconi}, M., {Molinaro}, R., {et~al.} 2020, \apjs, 247, 30

\bibitem[{{di Criscienzo} {et~al.}(2010){di Criscienzo}, {Ventura}, {D'Antona},
  {Milone}, \& {Piotto}}]{dicr10}
{di Criscienzo}, M., {Ventura}, P., {D'Antona}, F., {Milone}, A., \& {Piotto},
  G. 2010, \mnras, 408, 999

\bibitem[{{Doi} {et~al.}(2010){Doi}, {Tanaka}, {Fukugita}, {Gunn}, {Yasuda},
  {Ivezi{\'c}}, {Brinkmann}, {de Haars}, {Kleinman}, {Krzesinski}, \& {French
  Leger}}]{Doi10}
{Doi}, M., {Tanaka}, M., {Fukugita}, M., {et~al.} 2010, \aj, 139, 1628

\bibitem[{{Dotter} {et~al.}(2007){Dotter}, {Chaboyer}, {Ferguson}, {Lee},
  {Worthey}, {Jevremovi{\'c}}, \& {Baron}}]{dotter:07}
{Dotter}, A., {Chaboyer}, B., {Ferguson}, J.~W., {et~al.} 2007, \apj, 666, 403

\bibitem[{{Dotter} {et~al.}(2008{\natexlab{a}}){Dotter}, {Chaboyer},
  {Jevremovi{\'c}}, {Kostov}, {Baron}, \& {Ferguson}}]{dotter:08}
{Dotter}, A., {Chaboyer}, B., {Jevremovi{\'c}}, D., {et~al.}
  2008{\natexlab{a}}, \apjs, 178, 89

\bibitem[{{Dotter} {et~al.}(2008{\natexlab{b}}){Dotter}, {Chaboyer},
  {Jevremovi{\'c}}, {Kostov}, {Baron}, \& {Ferguson}}]{Dotter08}
---. 2008{\natexlab{b}}, \apjs, 178, 89

\bibitem[{{Fiorentino} {et~al.}(2006){Fiorentino}, {Limongi}, {Caputo}, \&
  {Marconi}}]{fiorentino06}
{Fiorentino}, G., {Limongi}, M., {Caputo}, F., \& {Marconi}, M. 2006, \aap,
  460, 155

\bibitem[{{Fiorentino} {et~al.}(2007){Fiorentino}, {Marconi}, {Musella}, \&
  {Caputo}}]{fiorentino07}
{Fiorentino}, G., {Marconi}, M., {Musella}, I., \& {Caputo}, F. 2007, \aap,
  476, 863

\bibitem[{{Fu} {et~al.}(2018){Fu}, {Bressan}, {Marigo}, {Girardi},
  {Montalb{\'a}n}, {Chen}, \& {Nanni}}]{parsecae}
{Fu}, X., {Bressan}, A., {Marigo}, P., {et~al.} 2018, \mnras, 476, 496

\bibitem[{{Gaia Collaboration} {et~al.}(2020){Gaia Collaboration}, {Brown},
  {Vallenari}, {Prusti}, {de Bruijne}, {Babusiaux}, \& {Biermann}}]{gaiadr3}
{Gaia Collaboration}, {Brown}, A.~G.~A., {Vallenari}, A., {et~al.} 2020, arXiv
  e-prints, arXiv:2012.01533

\bibitem[{{Girardi} {et~al.}(2002){Girardi}, {Bertelli}, {Bressan}, {Chiosi},
  {Groenewegen}, {Marigo}, {Salasnich}, \& {Weiss}}]{Girardi02}
{Girardi}, L., {Bertelli}, G., {Bressan}, A., {et~al.} 2002, \aap, 391, 195

\bibitem[{{Gratton} {et~al.}(2019){Gratton}, {Bragaglia}, {Carretta},
  {D'Orazi}, {Lucatello}, \& {Sollima}}]{gratrev:19}
{Gratton}, R., {Bragaglia}, A., {Carretta}, E., {et~al.} 2019, \aapr, 27, 8

\bibitem[{{Gratton} {et~al.}(2003){Gratton}, {Bragaglia}, {Carretta},
  {Clementini}, {Desidera}, {Grundahl}, \& {Lucatello}}]{gr03}
{Gratton}, R.~G., {Bragaglia}, A., {Carretta}, E., {et~al.} 2003, \aap, 408,
  529

\bibitem[{{Grevesse} \& {Sauval}(1998)}]{gs:98}
{Grevesse}, N., \& {Sauval}, A.~J. 1998, \ssr, 85, 161

\bibitem[{{Groenewegen}(2006)}]{Groenewegen06}
{Groenewegen}, M.~A.~T. 2006, \aap, 448, 181

\bibitem[{{Hauschildt} \& {Baron}(1999)}]{Hauschildt99}
{Hauschildt}, P.~H., \& {Baron}, E. 1999, Journal of Computational and Applied
  Mathematics, 109, 41

\bibitem[{{Hayes} {et~al.}(2018){Hayes}, {Majewski}, {Shetrone},
  {Fern{\'a}ndez-Alvar}, {Allende Prieto}, {Schuster}, {Carigi}, {Cunha},
  {Smith}, {Sobeck}, {Almeida}, {Beers}, {Carrera}, {Fern{\'a}ndez-Trincado},
  {Garc{\'\i}a-Hern{\'a}ndez}, {Geisler}, {Lane}, {Lucatello}, {Matthews},
  {Minniti}, {Nitschelm}, {Tang}, {Tissera}, \& {Zamora}}]{hayes}
{Hayes}, C.~R., {Majewski}, S.~R., {Shetrone}, M., {et~al.} 2018, \apj, 852, 49

\bibitem[{{Hidalgo} {et~al.}(2018){Hidalgo}, {Pietrinferni}, {Cassisi},
  {Salaris}, {Mucciarelli}, {Savino}, {Aparicio}, {Silva Aguirre}, \&
  {Verma}}]{bastiac:18}
{Hidalgo}, S.~L., {Pietrinferni}, A., {Cassisi}, S., {et~al.} 2018, \apj, 856,
  125

\bibitem[{{Husser} {et~al.}(2013){Husser}, {Wende-von Berg}, {Dreizler},
  {Homeier}, {Reiners}, {Barman}, \& {Hauschildt}}]{Husser13}
{Husser}, T.-O., {Wende-von Berg}, S., {Dreizler}, S., {et~al.} 2013, \aap,
  553, A6

\bibitem[{{Jordi} {et~al.}(2010){Jordi}, {Gebran}, {Carrasco}, {de Bruijne},
  {Voss}, {Fabricius}, {Knude}, {Vallenari}, {Kohley}, \& {Mora}}]{Jordi10}
{Jordi}, C., {Gebran}, M., {Carrasco}, J.~M., {et~al.} 2010, \aap, 523, A48

\bibitem[{{Kalirai} {et~al.}(2012){Kalirai}, {Richer}, {Anderson}, {Dotter},
  {Fahlman}, {Hansen}, {Hurley}, {King}, {Reitzel}, {Rich}, {Shara}, {Stetson},
  \& {Woodley}}]{kal}
{Kalirai}, J.~S., {Richer}, H.~B., {Anderson}, J., {et~al.} 2012, \aj, 143, 11

\bibitem[{{Kim} {et~al.}(2002){Kim}, {Demarque}, {Yi}, \& {Alexand
  er}}]{kim:02}
{Kim}, Y.-C., {Demarque}, P., {Yi}, S.~K., \& {Alexand er}, D.~R. 2002, \apjs,
  143, 499

\bibitem[{{Korn} {et~al.}(2007){Korn}, {Grundahl}, {Richard}, {Mashonkina},
  {Barklem}, {Collet}, {Gustafsson}, \& {Piskunov}}]{korn07}
{Korn}, A.~J., {Grundahl}, F., {Richard}, O., {et~al.} 2007, \apj, 671, 402

\bibitem[{{Kroupa} {et~al.}(1993){Kroupa}, {Tout}, \& {Gilmore}}]{kroupa}
{Kroupa}, P., {Tout}, C.~A., \& {Gilmore}, G. 1993, \mnras, 262, 545

\bibitem[{{Kurucz}(1970)}]{kurucz70}
{Kurucz}, R.~L. 1970, SAO Special Report, 309

\bibitem[{{Ma{\'{\i}}z Apell{\'a}niz}(2006)}]{Maiz-Apellaniz06}
{Ma{\'{\i}}z Apell{\'a}niz}, J. 2006, \aj, 131, 1184

\bibitem[{{Ma{\'\i}z Apell{\'a}niz}(2017)}]{Maiz-Apellaniz17}
{Ma{\'\i}z Apell{\'a}niz}, J. 2017, \aap, 608, L8

\bibitem[{{Ma{\'\i}z Apell{\'a}niz} \& {Weiler}(2018)}]{Maiz-Apellaniz18}
{Ma{\'\i}z Apell{\'a}niz}, J., \& {Weiler}, M. 2018, \aap, 619, A180

\bibitem[{{Marconi} {et~al.}(2020){Marconi}, {Molinaro}, {Ripepi}, {Leccia},
  {Musella}, {De Somma}, {Gatto}, \& {Moretti}}]{marconi20}
{Marconi}, M., {Molinaro}, R., {Ripepi}, V., {et~al.} 2020, arXiv e-prints,
  arXiv:2011.06675

\bibitem[{{Marconi} {et~al.}(2015){Marconi}, {Coppola}, {Bono}, {Braga},
  {Pietrinferni}, {Buonanno}, {Castellani}, {Musella}, {Ripepi}, \&
  {Stellingwerf}}]{marconi15}
{Marconi}, M., {Coppola}, G., {Bono}, G., {et~al.} 2015, \apj, 808, 50

\bibitem[{{Mashonkina} {et~al.}(2019){Mashonkina}, {Neretina}, {Sitnova}, \&
  {Pakhomov}}]{mashonkina}
{Mashonkina}, L.~I., {Neretina}, M.~D., {Sitnova}, T.~M., \& {Pakhomov}, Y.~V.
  2019, Astronomy Reports, 63, 726

\bibitem[{{Miglio} {et~al.}(2012){Miglio}, {Brogaard}, {Stello}, {Chaplin},
  {D'Antona}, {Montalb{\'a}n}, {Basu}, {Bressan}, {Grundahl}, {Pinsonneault},
  {Serenelli}, {Elsworth}, {Hekker}, {Kallinger}, {Mosser}, {Ventura},
  {Bonanno}, {Noels}, {Silva Aguirre}, {Szabo}, {Li}, {McCauliff}, {Middour},
  \& {Kjeldsen}}]{miglio:12}
{Miglio}, A., {Brogaard}, K., {Stello}, D., {et~al.} 2012, \mnras, 419, 2077

\bibitem[{{Milone} {et~al.}(2012){Milone}, {Marino}, {Cassisi}, {Piotto},
  {Bedin}, {Anderson}, {Allard}, {Aparicio}, {Bellini}, {Buonanno}, {Monelli},
  \& {Pietrinferni}}]{mil}
{Milone}, A.~P., {Marino}, A.~F., {Cassisi}, S., {et~al.} 2012, \apjl, 754, L34

\bibitem[{{Milone} {et~al.}(2018){Milone}, {Marino}, {Renzini}, {D'Antona},
  {Anderson}, {Barbuy}, {Bedin}, {Bellini}, {Brown}, {Cassisi}, {Cordoni},
  {Lagioia}, {Nardiello}, {Ortolani}, {Piotto}, {Sarajedini}, {Tailo}, {van der
  Marel}, \& {Vesperini}}]{mi18}
{Milone}, A.~P., {Marino}, A.~F., {Renzini}, A., {et~al.} 2018, \mnras, 481,
  5098

\bibitem[{{Mucciarelli} {et~al.}(2011){Mucciarelli}, {Salaris}, {Lovisi},
  {Ferraro}, {Lanzoni}, {Lucatello}, \& {Gratton}}]{mucc11}
{Mucciarelli}, A., {Salaris}, M., {Lovisi}, L., {et~al.} 2011, \mnras, 412, 81

\bibitem[{{Nieuwenhuijzen} \& {de Jager}(1990)}]{dejager}
{Nieuwenhuijzen}, H., \& {de Jager}, C. 1990, \aap, 231, 134

\bibitem[{{Pietrinferni} {et~al.}(2004){Pietrinferni}, {Cassisi}, {Salaris}, \&
  {Castelli}}]{basti:04}
{Pietrinferni}, A., {Cassisi}, S., {Salaris}, M., \& {Castelli}, F. 2004, \apj,
  612, 168

\bibitem[{{Pietrinferni} {et~al.}(2006){Pietrinferni}, {Cassisi}, {Salaris}, \&
  {Castelli}}]{basti:06}
---. 2006, \apj, 642, 797

\bibitem[{{Pietrinferni} {et~al.}(2013){Pietrinferni}, {Cassisi}, {Salaris}, \&
  {Hidalgo}}]{basti:13}
{Pietrinferni}, A., {Cassisi}, S., {Salaris}, M., \& {Hidalgo}, S. 2013, \aap,
  558, A46

\bibitem[{{Pietrinferni} {et~al.}(2009){Pietrinferni}, {Cassisi}, {Salaris},
  {Percival}, \& {Ferguson}}]{basti:09}
{Pietrinferni}, A., {Cassisi}, S., {Salaris}, M., {Percival}, S., \&
  {Ferguson}, J.~W. 2009, \apj, 697, 275

\bibitem[{{Poole} {et~al.}(2008){Poole}, {Breeveld}, {Page}, {Land sman},
  {Holland}, {Roming}, {Kuin}, {Brown}, {Gronwall}, {Hunsberger}, {Koch},
  {Mason}, {Schady}, {vanden Berk}, {Blustin}, {Boyd}, {Broos}, {Carter},
  {Chester}, {Cucchiara}, {Hancock}, {Huckle}, {Immler}, {Ivanushkina},
  {Kennedy}, {Marshall}, {Morgan}, {Pandey}, {de Pasquale}, {Smith}, \&
  {Still}}]{Poole08}
{Poole}, T.~S., {Breeveld}, A.~A., {Page}, M.~J., {et~al.} 2008, \mnras, 383,
  627

\bibitem[{{Ram{\'\i}rez} {et~al.}(2012){Ram{\'\i}rez}, {Mel{\'e}ndez}, \&
  {Chanam{\'e}}}]{melendez}
{Ram{\'\i}rez}, I., {Mel{\'e}ndez}, J., \& {Chanam{\'e}}, J. 2012, \apj, 757,
  164

\bibitem[{{Reimers}(1975)}]{reimers}
{Reimers}, D. 1975, Memoires of the Societe Royale des Sciences de Liege, 8,
  369

\bibitem[{{Richard} {et~al.}(2002){Richard}, {Michaud}, {Richer}, {Turcotte},
  {Turck-Chi{\`e}ze}, \& {VandenBerg}}]{richard}
{Richard}, O., {Michaud}, G., {Richer}, J., {et~al.} 2002, \apj, 568, 979

\bibitem[{{Richer} {et~al.}(2008){Richer}, {Dotter}, {Hurley}, {Anderson},
  {King}, {Davis}, {Fahlman}, {Hansen}, {Kalirai}, {Paust}, {Rich}, \&
  {Shara}}]{richer08}
{Richer}, H.~B., {Dotter}, A., {Hurley}, J., {et~al.} 2008, \aj, 135, 2141

\bibitem[{{Ripepi} {et~al.}(2019){Ripepi}, {Molinaro}, {Musella}, {Marconi},
  {Leccia}, \& {Eyer}}]{ripepi19}
{Ripepi}, V., {Molinaro}, R., {Musella}, I., {et~al.} 2019, \aap, 625, A14

\bibitem[{{Rubele} {et~al.}(2012){Rubele}, {Kerber}, {Girardi}, {Cioni},
  {Marigo}, {Zaggia}, {Bekki}, {de Grijs}, {Emerson}, {Groenewegen},
  {Gullieuszik}, {Ivanov}, {Miszalski}, {Oliveira}, {Tatton}, \& {van
  Loon}}]{Rubele12}
{Rubele}, S., {Kerber}, L., {Girardi}, L., {et~al.} 2012, \aap, 537, A106

\bibitem[{{Salaris} \& {Cassisi}(2017)}]{screview:17}
{Salaris}, M., \& {Cassisi}, S. 2017, Royal Society Open Science, 4, 170192

\bibitem[{{Salaris} {et~al.}(2016){Salaris}, {Cassisi}, \&
  {Pietrinferni}}]{scp16}
{Salaris}, M., {Cassisi}, S., \& {Pietrinferni}, A. 2016, \aap, 590, A64

\bibitem[{{Salaris} {et~al.}(2010){Salaris}, {Cassisi}, {Pietrinferni},
  {Kowalski}, \& {Isern}}]{bastiwd}
{Salaris}, M., {Cassisi}, S., {Pietrinferni}, A., {Kowalski}, P.~M., \&
  {Isern}, J. 2010, \apj, 716, 1241

\bibitem[{{Salaris} {et~al.}(1993){Salaris}, {Chieffi}, \&
  {Straniero}}]{salaris:93}
{Salaris}, M., {Chieffi}, A., \& {Straniero}, O. 1993, \apj, 414, 580

\bibitem[{{Salaris} \& {Weiss}(1998)}]{sw:98}
{Salaris}, M., \& {Weiss}, A. 1998, \aap, 335, 943

\bibitem[{{Salaris} {et~al.}(2006){Salaris}, {Weiss}, {Ferguson}, \&
  {Fusilier}}]{swff}
{Salaris}, M., {Weiss}, A., {Ferguson}, J.~W., \& {Fusilier}, D.~J. 2006, \apj,
  645, 1131

\bibitem[{{Salasnich} {et~al.}(2000){Salasnich}, {Girardi}, {Weiss}, \&
  {Chiosi}}]{salasnich:00}
{Salasnich}, B., {Girardi}, L., {Weiss}, A., \& {Chiosi}, C. 2000, \aap, 361,
  1023

\bibitem[{{Schr{\"o}der} \& {Cuntz}(2005)}]{cuntz}
{Schr{\"o}der}, K.~P., \& {Cuntz}, M. 2005, \apjl, 630, L73

\bibitem[{{Stetson}(2000)}]{st00}
{Stetson}, P.~B. 2000, \pasp, 112, 925

\bibitem[{{Tandon} {et~al.}(2017){Tandon}, {Hutchings}, {Ghosh}, {Subramaniam},
  {Koshy}, {Girish}, {Kamath}, {Kathiravan}, {Kumar}, {Lancelot}, {Mahesh},
  {Mohan}, {Murthy}, {Nagabhushana}, {Pati}, {Postma}, {Rao},
  {Sankarasubramanian}, {Sreekumar}, {Sriram}, {Stalin}, {Sutaria}, {Sreedhar},
  {Barve}, {Mondal}, \& {Sahu}}]{Tandon17}
{Tandon}, S.~N., {Hutchings}, J.~B., {Ghosh}, S.~K., {et~al.} 2017, Journal of
  Astrophysics and Astronomy, 38, 28

\bibitem[{{Thompson} {et~al.}(2010){Thompson}, {Kaluzny}, {Rucinski},
  {Krzeminski}, {Pych}, {Dotter}, \& {Burley}}]{eb10}
{Thompson}, I.~B., {Kaluzny}, J., {Rucinski}, S.~M., {et~al.} 2010, \aj, 139,
  329

\bibitem[{{Thompson} {et~al.}(2020){Thompson}, {Udalski}, {Dotter}, {Rozyczka},
  {Schwarzenberg-Czerny}, {Pych}, {Beletsky}, {Burley}, {Marshall},
  {McWilliam}, {Morrell}, {Osip}, {Monson}, {Persson}, {Szyma{\'n}ski},
  {Soszy{\'n}ski}, {Poleski}, {Ulaczyk}, {Wyrzykowski}, {Koz{\l}owski},
  {Mr{\'o}z}, {Pietrukowicz}, \& {Skowron}}]{eb20}
{Thompson}, I.~B., {Udalski}, A., {Dotter}, A., {et~al.} 2020, \mnras, 492,
  4254

\bibitem[{{Tonry} {et~al.}(2012){Tonry}, {Stubbs}, {Lykke}, {Doherty},
  {Shivvers}, {Burgett}, {Chambers}, {Hodapp}, {Kaiser}, {Kudritzki},
  {Magnier}, {Morgan}, {Price}, \& {Wainscoat}}]{Tonry12}
{Tonry}, J.~L., {Stubbs}, C.~W., {Lykke}, K.~R., {et~al.} 2012, \apj, 750, 99

\bibitem[{{Turcotte} {et~al.}(1998){Turcotte}, {Richer}, {Michaud}, {Iglesias},
  \& {Rogers}}]{turcotte}
{Turcotte}, S., {Richer}, J., {Michaud}, G., {Iglesias}, C.~A., \& {Rogers},
  F.~J. 1998, \apj, 504, 539

\bibitem[{{Valcarce} {et~al.}(2012){Valcarce}, {Catelan}, \&
  {Sweigart}}]{valcarce}
{Valcarce}, A.~A.~R., {Catelan}, M., \& {Sweigart}, A.~V. 2012, \aap, 547, A5

\bibitem[{{VandenBerg} {et~al.}(2012){VandenBerg}, {Bergbusch}, {Dotter},
  {Ferguson}, {Michaud}, {Richer}, \& {Proffitt}}]{don:12}
{VandenBerg}, D.~A., {Bergbusch}, P.~A., {Dotter}, A., {et~al.} 2012, \apj,
  755, 15

\bibitem[{{VandenBerg} {et~al.}(2014){VandenBerg}, {Bergbusch}, {Ferguson}, \&
  {Edvardsson}}]{don:14}
{VandenBerg}, D.~A., {Bergbusch}, P.~A., {Ferguson}, J.~W., \& {Edvardsson}, B.
  2014, \apj, 794, 72

\bibitem[{{VandenBerg} {et~al.}(2000){VandenBerg}, {Swenson}, {Rogers},
  {Iglesias}, \& {Alexander}}]{don:00}
{VandenBerg}, D.~A., {Swenson}, F.~J., {Rogers}, F.~J., {Iglesias}, C.~A., \&
  {Alexander}, D.~R. 2000, \apj, 532, 430

\bibitem[{{Woo} {et~al.}(2003){Woo}, {Gallart}, {Demarque}, {Yi}, \&
  {Zoccali}}]{woo03}
{Woo}, J.-H., {Gallart}, C., {Demarque}, P., {Yi}, S., \& {Zoccali}, M. 2003,
  \aj, 125, 754

\bibitem[{{Wright} {et~al.}(2010){Wright}, {Eisenhardt}, {Mainzer}, {Ressler},
  {Cutri}, {Jarrett}, {Kirkpatrick}, {Padgett}, {McMillan}, {Skrutskie},
  {Stanford}, {Cohen}, {Walker}, {Mather}, {Leisawitz}, {Gautier}, {McLean},
  {Benford}, {Lonsdale}, {Blain}, {Mendez}, {Irace}, {Duval}, {Liu}, {Royer},
  {Heinrichsen}, {Howard}, {Shannon}, {Kendall}, {Walsh}, {Larsen}, {Cardon},
  {Schick}, {Schwalm}, {Abid}, {Fabinsky}, {Naes}, \& {Tsai}}]{Wright10}
{Wright}, E.~L., {Eisenhardt}, P. R.~M., {Mainzer}, A.~K., {et~al.} 2010, \aj,
  140, 1868

\end{thebibliography}

\appendix

\section{Synthetic color-magnitude diagram tool at BaSTI web site.}

As for the previous release of the library, the new BaSTI website contains a tool for the computation of synthetic CMDs (http://basti-iac.oa-teramo.inaf.it/syncmd.html). This tool
can be used after requesting a user ID to the BaSTI-IAC team members by using the link http://basti-iac.oa-abruzzo.inaf.it/requests.html.
Here we provide information about the inputs
and how the code works. The user has to select among a combination of heavy element mixtures (solar scaled or
$\alpha$-enhanced) and
available grids of models (various options about overshooting, diffusion and mass loss).
Variations of the He abundance at fixed metallicity cannot yet be considered, but this
  is a feature that will be implemented in the near future.
The user can also choose to identify the radial pulsators in the synthetic population, and determine their type and pulsation periods.

 After this selection, two sets 
of input parameters are requested: Star-formation history (SFH) and photometric input parameters. SFH input parameters
are as follows: 

\begin{itemize}
\item {\bf Age}: A list of ages $t_i$ (in Myr, older age first, with age = 0 denoting stars that are forming now). A maximum of 50
  age values are allowed.
\item {\bf SFR}: Relative star formation rate at each age. The code rescales the individual values to the maximum
  one provided.
  \item {\bf Metallicity}: [Fe/H] of stars formed at each $t_i$.
  \item {\bf Metallicity spread}: 1 sigma Gaussian spread (in dex) around each metallicity. 
  \item {\bf SFR scale}: This number (the maximum value allowed is equal to  $2\times 10^6$)
    is multiplied by the value of the {\bf SFR} to provide the number of stars formed between
    age $t_i$ and age $t_{i+1}$.
  \item {\bf Low mass}: Lower stellar mass mass (in units of $\rm M_\odot$) to be included in the calculations (between
    $0.1~\rm M_\odot$ and $120~\rm M_\odot$).
  \item {\bf Binaries}: Fraction of unresolved binaries. If the fraction is different from zero
    the mass of the second component is
    selected randomly following \citet{woo03}, and the fluxes of
    the two unresolved components are properly added.
  \item {\bf Mass ratio}: Minimum mass ratio for binary systems (upper value is 1.0).
  \item {\bf IMF}: Initial mass function type (0 for a single power law, 1 for \citet{kroupa}).
    If a single power law is selected, the slope must be given (e.g.: -2.35).
  \item {\bf Variables}: If a value equal to 1 is assigned to this parameter, 
    the code identifies the radial pulsators in the synthetic population, and calculates their  
    properties. A value equal to 0 makes the code skip the identification of radial pulsators.
  \item {\bf Random 1 and 2}: Seeds for the Monte Carlo number generator.
    The system will generate these numbers automatically if none are given.
\end{itemize}

The photometric input parameters are:

\begin{itemize}
  \item {\bf Photometric error}: Mean photometric error (mag.) 
  \item {\bf Photometric error type}: None, constant, or error table.
  \item {\bf Photometric system}: Select one of the photometric systems available.
\end{itemize}

\begin{figure}[ht!]
\begin{center}
\includegraphics[width=3.0in]{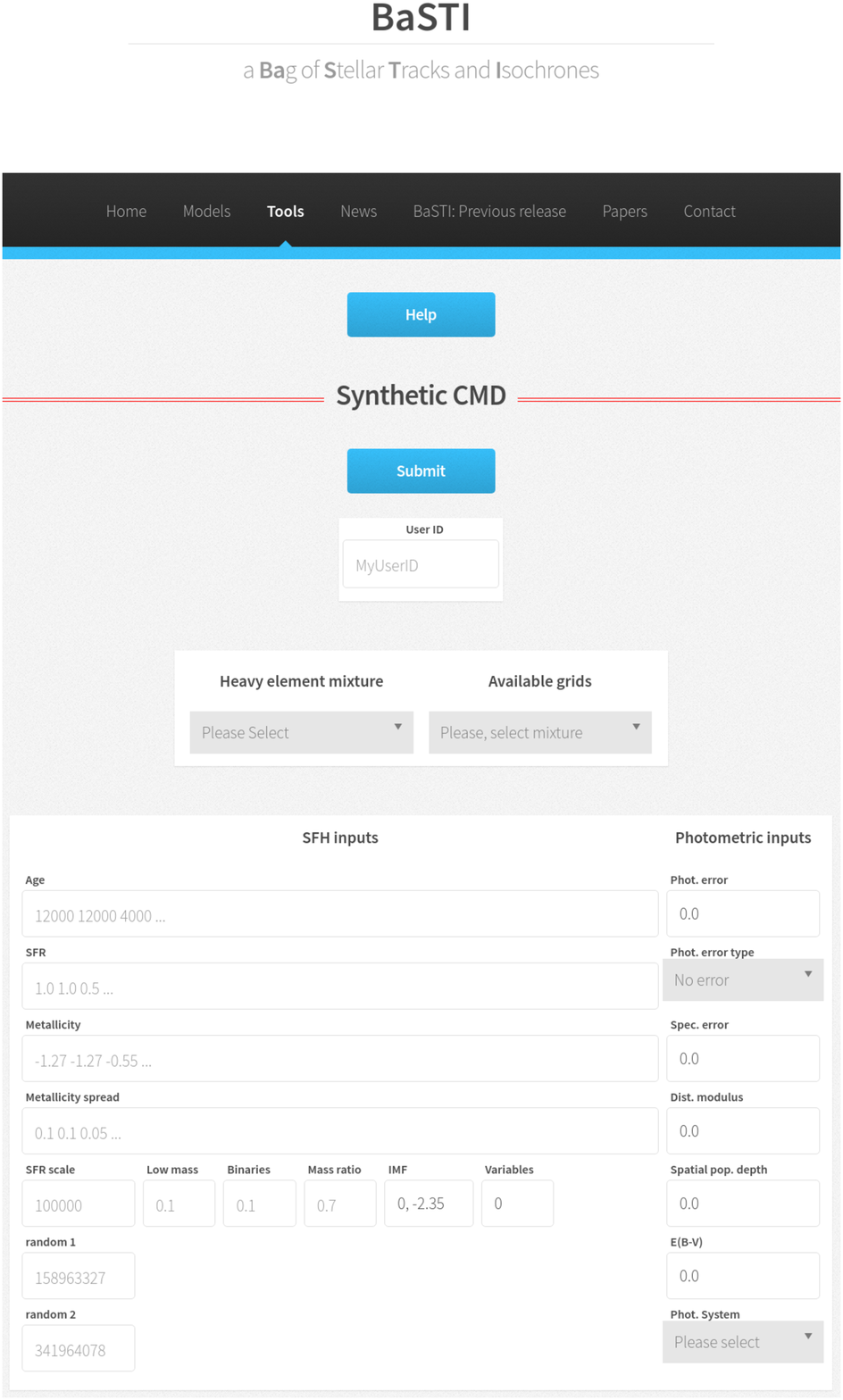}
\caption{Synthetic CMD web-tool at BaSTI website.\label{fig:webtool}}
\end{center}
\end{figure}

The code computes the synthetic CMD as follows: For each age $t_i$, the number
of stars formed between $t_i$ and $t_{i+1}$ is obtained by the multiplying the SFR scale by the
value of the SFR at $t_i$. For each star born in this age interval, a random age t ($t_i \le t < t_{i+1}$) is drawn from a flat
probability distribution, together with a mass $m$ selected according to the specified IMF, and the corresponding value of
$\rm [Fe/H]_i$. If a value different from zero for $\rm \sigma([Fe/H]_i)$ is specified, then $\rm [Fe/H]_i$is perturbed using a
Gaussian probability distribution centered in $\rm [Fe/H]_i$ and sigma = $\rm \sigma([Fe/H]_i)$. With these three values of $t$,
$m$, and [Fe/H] the program interpolates quadratically in age, metallicity and mass among the isochrones in the grid, to calculate
the star's photometric properties, plus luminosity and effective temperature.
The code also checks whether the synthetic star is located within the instability strip (IS)
  for radial pulsations, by comparing its position in the HRD with the boundaries of the IS
  predicted by accurate pulsation models of RR Lyrae stars \citep[see,][and references therein]{marconi15}, anomalous Cepheids
  \citep[][]{fiorentino06}, and classical Cepheids \citep[][]{fiorentino07,desomma20}. If the star
  is located within a given IS, the corresponding pulsation period is calculated by using the appropriate theoretical 
  relationship (see the previous references) between period, luminosity, effective temperature, mass and metallicity.

Once all stars formed between ages $t_i$ and $t_{i+1}$ are generated, the next time interval is
considered and the cycle is repeated, ending when all stars in the final age bin between $t_{n-1}$ and $t_n$ are generated. The values
of the SFR and [Fe/H] at $t_n$ are not considered and can be set to any arbitrary number. To compute the synthetic CMD of a
single-age stellar population, just one age value needs to be provided as input. The BaSTI website provides some examples of
SFHs and the corresponding web-tool inputs.

Once a run is completed, the user will receive an email with instructions to download two files: One with the synthetic stars, and
another with the integrated properties of the population. The content of the first file is as follows:

\begin{itemize}
  \item column 1: Star number (+2 if unresolved binary);
  \item column 2: Logarithm of the age in years;
  \item column 3: [Fe/H];                              
  \item column 4: Value of the current stellar mass in $\rm M_\odot$;       
  \item column 5: log($L/L_\odot$);                           
  \item column 6: log($T_{\mathrm eff}$);                           
  \item column 7: Initial mass of the unresolved secondary star ($\rm M_\odot$) if different from 0.0; 
  \item column 8: Index that denotes the type of radial pulsator. A value equal to
    0 stands for no pulsations, 1 corresponds to fundamental-mode RR Lyraes,
    2 identifies the first overtone RR Lyraes,  3 corresponds to
    fundamental-mode anomalous Cepheids, 4 labels the first overtone anomalous Cepheids,; 5
    denotes fundamental-mode classical Cepheids;
  \item column 9: log(P), with $P$ being the period of pulsations (in days). It is
    set to 99.99 if the synthetic star does not pulsate (see previous discussion);
  \item column 10 to the end: Absolute magnitudes in the selected photometric system.
\end{itemize}

The integrated properties file contains the following information:

\begin{itemize}
  \item Integrated magnitudes in all bands for the selected photometric system.
  \item Total mass of formed stars ($\rm M_\odot$).
  \item Number of stars evolving in the synthetic CMD, including unresolved stars companions.
  \item Number of fundamental-mode RR-Lyrae stars.
  \item Number of first overtone RR-Lyrae stars.
  \item Number of fundamental-mode anomalous Cepheids.
  \item Number of first overtone anomalous Cepheids.
  \item Number of fundamental-mode classical Cepheids
\end{itemize}

\end{document}